\documentclass[prd,amsfonts,twocolumn,nofootinbib,showpacs,dvipdfmx]{revtex4}
\RequirePackage{graphicx}
\usepackage{enumerate}
\usepackage{bm}
\usepackage{braket}
\usepackage{slashed}
\usepackage{color}
\usepackage{comment}
\usepackage{ulem}
\begin{document}
\title{The Exact WKB and the Landau-Zener transition for
asymmetry in cosmological particle production}

\author{Seishi Enomoto}
\affiliation{School of Physics, Sun Yat-sen University, Guangzhou 510275, China}
\author{Tomohiro Matsuda}
\affiliation{Laboratory of Physics, Saitama Institute of Technology,
Fukaya, Saitama 369-0293, Japan}
\begin{abstract}
Cosmological particle production by a time-dependent scalar field is
common in cosmology.
We focus on the mechanism of asymmetry production when
interaction explicitly violates symmetry and its motion is rapid enough to
 create particles by itself.
Combining the exact WKB analysis and the Landau-Zener transition, 
we point out that perturbation before the non-perturbative analysis
 may drastically change the structure of the Stokes lines of the theory.
The Exact WKB can play an important role in avoiding such discrepancies.
\end{abstract}

\maketitle
\section{Introduction}
In 1991, Cohen, Kaplan and Nelson proposed a mechanism of ``Spontaneous
Baryogenesis'' for producing baryons at the electroweak phase transition
in the adiabatic limit of thick slowly moving bubble walls\cite{Cohen:1991iu}.
Their original idea\cite{Cohen:1987vi,Cohen:1988kt} uses an effective
chemical potential for biasing the baryon number, and the effective
chemical potential was brought in by considering a time-dependent parameter.
The mechanism avoids the ``out of thermal equilibrium'' condition in the
famous Sakharov's three conditions\cite{Sakharov:1967dj} since the
time-dependent background violates CPT.
The mechanism has been considered in many models of
baryogenesis since
it has been obvious that the mechanism is quite useful for constructing
 mechanisms for generating the baryon number of the Universe.
On the other hand, it has been suggested that the effective chemical 
potential may disappear from the Hamiltonian formalism
when the field equation of
the time-dependent parameter is taken into account\cite{Arbuzova:2016qfh}.
For us, this point is one of the primary reasons for considering
(complex) fundamental
equations, instead of using the (useful) effective theory.
In this paper, we are not considering thermal equilibrium, but the basic
idea relies on the spontaneous baryogenesis scenario.

In past studies, such as Ref.\cite{Pearce:2015nga,
Adshead:2015jza,Adshead:2015kza}, 
baryogenesis with non-perturbative particle production has been
discussed with a chemical potential.
To show clearly the purpose of this paper,
we first explain how ``chemical potential'' affects the non-perturbative
particle production.

Let us start with the simplest scenario of bosonic preheating given by
the action\cite{Enomoto:2017rvc}
\begin{eqnarray}
S_0&=&\int d^4 x\sqrt{-g}\left[\partial_\mu\phi^*\partial^{\mu}\phi
-m^2 |\phi|^2+\xi R|\phi|^2 \right].
\end{eqnarray}
Using conformal time $\eta$, one can write the metric
$g_{\mu\nu}=a^2(\eta){\rm diag}(1,-1,-1,-1)$ and $R=-6\ddot{a}/a^3$,
where $a$ is the cosmological scale factor and the dot denotes
time-derivative with respect to the conformal time.
A convenient definition of a new field is $\chi\equiv a\phi$, which
gives a simple form
\begin{eqnarray}
S_0&=&\int d^4 x \left[|\dot{\chi}|^2
-\omega^2 |\chi|^2\right],
\end{eqnarray}
where
\begin{eqnarray}
\omega^2&\equiv& a^2m^2 +
\left(-\Delta + \frac{\ddot{a}}{a}(6\xi-1)\right).
\end{eqnarray}
Here $\Delta$ is the Laplacian.
Annihilation ($a,b$) and creation ($a^\dagger,b^\dagger$) operators of
``particle'' and ``antiparticle'' appear in the decomposition
\begin{eqnarray}
\chi&=& \int \frac{d^3 k}{(2\pi)^{3/2}}\left[
h(\eta) a(\bm{k}) e^{i \bm{k}\cdot \bm{x}}
+g^*(\eta) b^\dagger(\bm{k}) e^{-i \bm{k}\cdot \bm{x}}\right].\nonumber\\
\end{eqnarray}
For our calculation, we introduce conjugate momenta
$\Pi^\dagger\equiv\dot{\chi}$,
which can be decomposed as
\begin{eqnarray}
\Pi^\dagger&=& \int \frac{d^3 k}{(2\pi)^{3/2}}\left[
\tilde{h}(\eta) a(\bm{k}) e^{i \bm{k}\cdot \bm{x}}
+\tilde{g}^*(\eta) b^\dagger(\bm{k}) e^{-i \bm{k}\cdot \bm{x}}\right].\nonumber\\
\end{eqnarray}
Following Ref.\cite{ZS-original}, we expand
$h, \tilde{h}$ (particles) and $g, \tilde{g}$ (antiparticles) as 
\begin{eqnarray}
h&=&\frac{e^{-i\int^\eta \omega d\eta'}}{\sqrt{2\omega}}A_h
+\frac{e^{i\int^\eta \omega d\eta'}}{\sqrt{2\omega}}B_h,\nonumber\\
\tilde{h}&=&\frac{-i\omega e^{-i\int^\eta \omega d\eta'}}{\sqrt{2\omega}}A_h
+\frac{i\omega e^{i\int^\eta \omega d\eta'}}{\sqrt{2\omega}}B_h,
\end{eqnarray}
and
\begin{eqnarray}
g&=&\frac{e^{-i\int^\eta \omega d\eta'}}{\sqrt{2\omega}}A_g
+\frac{e^{i\int^\eta \omega d\eta'}}{\sqrt{2\omega}}B_g,\nonumber\\
\tilde{g}&=&\frac{-i\omega e^{-i\int^\eta \omega d\eta'}}{\sqrt{2\omega}}A_g
+\frac{i\omega e^{i\int^\eta \omega d\eta'}}{\sqrt{2\omega}}B_g,
\end{eqnarray}
where $A$ and $B$ are known as the Bogoliubov coefficients.
For further simplification, we introduce $\alpha$ and $\beta$, which are
defined as
\begin{eqnarray}
\alpha_{h,g}&\equiv& e^{-i\int^\eta\omega d\eta'}A_{h,g}\\
\beta_{h,g}&\equiv& e^{i\int^\eta\omega d\eta'}B_{h,g}.
\end{eqnarray}
Now the equation of motion can be written as
\begin{eqnarray}
\dot{h}-\tilde{h}&=&0\\
\dot{\tilde{h}}+\omega^2 h&=&0,
\end{eqnarray}
which are solved for $\dot{\alpha}$ and $\dot{\beta}$ as 
\begin{eqnarray}
\dot{\alpha}_h&=&-i\omega \alpha_h
 +\frac{\dot{\omega}}{2\omega}\beta_h\nonumber\\
\dot{\beta}_h&=&i\omega \beta_h
 +\frac{\dot{\omega}}{2\omega}\alpha_h.
\end{eqnarray}

Let us see what happens when a constant chemical potential is introduced.
After adding a chemical potential
\begin{eqnarray}
{\cal L}&=&\dot{\chi}\dot{\chi}^* -\omega^2 |\chi|^2
-i\mu_\chi \left(\chi \dot{\chi}^*-\chi^* \dot{\chi}\right),
\end{eqnarray}
we find
\begin{eqnarray}
\label{eq-of-mo-boson}
\ddot{\chi}-2i\mu_\chi\dot{\chi}+(\omega^2-i\dot{\mu}_\chi)\chi&=&0.
\end{eqnarray}
There are two terms which might cause differences.
One is $-2i\mu_\chi\dot{\chi}$, and the other is $-i\dot{\mu}_\chi \chi$.
If one assumes a constant chemical potential, only the first term will
remain. 
{\bf Rather surprisingly, a constant chemical potential does not generate
asymmetry.
The reason will become very clear when the EWKB formalism is introduced,
but here we will follow the standard formalism.}
Then the equation of motion can be written as
\begin{eqnarray}
\dot{h}-\tilde{h}-i\mu_\chi h&=&0\nonumber\\
\dot{\tilde{h}}+\omega^2h -i\mu_\chi\tilde{h}&=&0,
\end{eqnarray}
where a complex parameter ($\sim i\mu_\chi$) appears.
One can solve these equations for $\dot{\alpha}$ and $\dot{\beta}$ to
find
\begin{eqnarray}
\dot{\alpha}_h&=&-i(\omega-\mu_\chi)\alpha_h
 +\frac{\dot{\omega}}{2\omega}\beta_h\nonumber\\
\dot{\beta}_h&=&\frac{\dot{\omega}}{2\omega}\alpha_h+i(\omega+\mu_\chi)\beta_h.
\end{eqnarray}
and
\begin{eqnarray}
\dot{\alpha}_g&=&-i(\omega+\mu_\chi)\alpha_g
 +\frac{\dot{\omega}}{2\omega}\beta_g\nonumber\\
\dot{\beta}_g&=&\frac{\dot{\omega}}{2\omega}\alpha_g+i(\omega-\mu_\chi)\beta_g.
\end{eqnarray}
One could naively claim that the shift of $\omega\pm
\mu_\chi$ is the source of the asymmetry.
{\bf 
However, this naive speculation fails in the present model.}
One can calculate the behavior of $|\beta|^2$  (both numerically and
analytically\cite{Enomoto:2017rvc}\footnote{The ``constant'' chemical
potential just affects the phase rotation of $\alpha_{h.g}$ and
$\beta_{h,g}$,
and thus it does not appear in the physical quantity. Indeed, one can
easily find that the equation of
motion for $|\beta_{h,g}|^2$ does not depend on $\mu_\chi$.}) to find
that the evolution of $|\beta_h|^2$ and $|\beta_g|^2$ are identical in
this case, resulting no asymmetry production.
From this simple model, one can understand why $\dot{\mu}\ne 0$ (i.e, a
time-dependent chemical potential) is needed for the asymmetry
production.

Using the simplest model, we have seen that a constant chemical
potential may not source the asymmetry.
Although the result may depend on the details of the model, 
what is important here is that the meaning of ``chemical potential''
is becoming vague for the non-perturbative particle production scenario.
This is why we have introduced mathematical tools for analyzing the
asymmetry.
Of course, in reality the above scenario should be considered with a
time-dependent chemical potential, 
since usually $\mu$ is defined using a time-dependent parameter and such
parameter is normally time-dependent during cosmological evolution.
Therefore, usually the numerical calculation of a phenomenological model
will generate the asymmetry, but still the meaning of ``chemical
potential'' is vague.

On the other hand, if the chemical potential is considered for a system of
Boltzmann equations, the complexities discussed above for the
non-perturbative particle production will not appear.
In this sense, arguments of the chemical potential have to be
discriminated between non-perturbative particle production and a system
of Boltzmann equations.
See also the recent arguments on the Higgs relaxation in
Ref.\cite{Kusenko:2014lra,Yang:2015ida,Wu:2019ohx}.

In this paper, we analyze cosmological particle production by a
time-dependent interaction.
We analytically explain the reason and the requirements of
asymmetric particle production in typical situations.
To avoid confusion, here we note that
normally such ``asymmetric particle production'' is explained by two 
stages; (symmetric) production of heavy particles and asymmetric decay
of the heavy particles, where the asymmetry is usually due to the interference.
In this sense, our strategy is not common, as we are considering
direct asymmetry production from the time-dependent scalar field.
Although not very common, direct asymmetry production has a
long history.
Dolgov et. al.\cite{Dolgov:1994zq, Dolgov:1996qq} calculated
the baryon asymmetry created by the decay of a pseudo-Nambu-Goldstone
boson (PNGB), whose interactions violate baryon number conservation of
fermions. 
Their calculation of Ref.\cite{Dolgov:1996qq} considers the Bogoliubov
transformation after perturbative expansion.
We are considering their calculation as a reference model.
Compared with their calculation, our calculation is rather technical.
Differences will be clearly described in this paper.

For scalar fields, Funakubo et. al.\cite{Funakubo:2000us} and 
Rangarajan and Nanopoulos\cite{Rangarajan:2001yu} calculated asymmetric
particle production.
See also recent developments in this direction in
Refs.\cite{Kusenko:2014uta, Adshead:2015jza, Adshead:2015kza, Enomoto:2017rvc,
Enomoto:2018yeu, Enomoto:2020lpf}. 

The original scenario of spontaneous baryogenesis uses the rather
moderate motion of a background field to source the effective chemical
potential in the thermal background.
On the other hand, our focus in this paper is rapid motion, which
(itself) can cause efficient particle production.
In this direction, the most famous scenario in cosmology would be the preheating
scenario, which discusses non-perturbative particle production
before reheating\cite{Dolgov:1989us, Kofman:1997yn}.
Besides the preheating scenario, there are many papers considering the
famous Schwinger mechanisms in cosmology.
The Schwinger mechanism\cite{Schwinger:1951nm}, which is named after
Schwinger who first derived the exponential formula for the pair
production, is still an
active research target\cite{Shakeri:2019mnt, Kitamoto:2020tjm, Taya:2020dco}. 
We also suppose phase transition or decay of
unstable domain wall networks for our scenario, which is also expected
to cause similar particle production. 
Since the configuration of scalar fields during the evolution of the
Universe may develop domain wall
structure and such configuration has to decay before nucleosynthesis, 
it would be interesting if decaying domain walls can generate baryon
numbers.\footnote{A natural mechanism of generating
safe(unstable) domain walls in supersymmetric theory has been advocated
in Ref.\cite{Matsuda:1998ms}.
See also Ref.\cite{Dolgov:2015gqa} for matter-antimatter asymmetry and
safe domain walls.}
Since the conventional $Z_n$-domain wall is interpolating between vacua
with different phases, the phase of the field becomes the primary time-dependent
parameter in such a scenario.
Besides the particle production, the scattering of fermions by the
walls could be asymmetric\cite{Nelson:1991ab, Funakubo:1996gi}.
This idea has been used for baryogenesis at the electroweak
scale\cite{Nelson:1991ab}.

Although there are many scenarios of cosmology in which asymmetric
particle production could be important, we will not discuss
details of the phenomenological aspects and focus on the technical
aspects of asymmetry production. 

To avoid confusion, we first explain the crucial difference between the
conventional preheating scenarios and our approach.
Since the ``symmetry violating interaction'' inevitably requires
multiple fields, our original equations have to be multicomponent differential
equations.
Although the typical single-field equation of the conventional preheating
scenario can be solved using the special function, it is impossible
to obtain such a solution in general.
Therefore, we need to develop mathematical methods to get an
analytical estimation of the asymmetry generated from the equations.
This includes sensible approximations and methods of calculating transfer
matrix between asymptotic solutions when the exact solutions are not
 written by the special function.
To avoid this problem, previous approaches\cite{Dolgov:1989us,
Funakubo:2000us, Rangarajan:2001yu} sometimes use perturbative
expansion before the non-perturbative analysis, where 
the special function can be used for the unperturbed solution.
However, such expansion may drastically change the structure of the
Stokes lines of the original theory.
To avoid the problem, one has to understand the Stokes lines of the
original theory first.
Some concrete examples will be shown in this paper.

In this paper, we consider the Landau-Zener model and the 
Exact WKB analysis (EWKB) for understanding the Stokes phenomena of the
particle creation\footnote{In
Ref.\cite{Enomoto:2020xlf}, we have applied the EWKB to cosmological
particle production (without asymmetry). See Ref.\cite{Enomoto:2020xlf,
Sueishi:2020rug, Taya:2020dco, Sueishi:2021xti} for more references.}.
As we will show in this paper, the combination of these methods is very
useful in understanding the origin of the asymmetry.
Theoretically, the extension of the EWKB calculation to a higher Landau-Zener
model is straightforward\cite{Virtual:2015HKT}, but because of the complexity of
the analytical result (it contains solutions of higher-order equation), we
are reducing the equations to the conventional two-component model.
For multiple Dirac fermions, we are taking the relativistic 
and the non-relativistic limits.

For fermions, our equations can be regarded as a generalized
Landau-Zener model\cite{Zener:1932ws}.
This analogy is sometimes very useful for understanding the origin of
the asymmetry.
Although the original Landau-Zener model mainly considers time-dependent
diagonal elements, our focus is the rotational motion of the
off-diagonal elements.
We consider such models since the off-diagonal elements are supposed to
be coming from the required interaction (i.e, symmetry violation)
of asymmetry production.
Mathematically, the time-dependence of the off-diagonal elements can be moved into the
diagonal elements using some transformation.

To explain the basic ideas of our strategy, we start with the
solution of the original Landau-Zener model in the next section, in
which transition between states is calculated when diagonal elements
are time-dependent.
Since the ``adiabatic states'' are diagonalizing the
Hamiltonian, particle production can be calculated from the Landau-Zener
transition, which is very convenient.
In the next section, it will be clear why the transition matrix of the
Landau-Zener model explains the Bogoliubov transformation of the
cosmological particle production.

See appendix \ref{app-srevEWKB} and \ref{app-reviewEWKB} for more technical
details of the EWKB
and Landau-Zener transformation applied to cosmological particle production.

\subsection{The Landau-Zener model and particle creation in cosmology}
First, we review the original Landau-Zener model and explain how it
can be related to cosmological particle production.
Here, the ``velocity'' is $v>0$, and the off-diagonal element $\Delta$
is supposed to be real.
The Landau-Zener model uses a couple of ordinary differential equations
given by
\begin{eqnarray}
i\hbar\frac{d}{dt}\left(
\begin{array}{c}
\psi_1\\
\psi_2
\end{array}
\right)&=&\left(
\begin{array}{cc}
-\frac{v}{2}t& \Delta \\
 \Delta& +\frac{v}{2}t 
\end{array}
\right)
\left(
\begin{array}{c}
\psi_1\\
\psi_2
\end{array}
\right),
\end{eqnarray}
which can be decoupled to give 
\begin{eqnarray}
\left[\hbar^2\frac{d^2}{dt^2}+\left(\Delta^2-i\hbar\frac{v}{2}\right)+\frac{1}{4}v^2t^2\right]\psi_1&=&0\\
\left[\hbar^2\frac{d^2}{dt^2}+\left(\Delta^2+i\hbar\frac{v}{2}\right)+\frac{1}{4}v^2t^2\right]\psi_2&=&0.
\end{eqnarray}
Following Ref.\cite{Virtual:2015HKT, EWKB}, we are going to rewrite the
 equations in the standard EWKB form.
In this form, a ``large'' parameter $\eta\equiv \hbar^{-1}$ is
introduced to give the ``Schr\"odinger equation'' 
\begin{eqnarray}
\left[-\frac{d^2}{dx^2}+\eta^2 Q(x)
\right]\psi(x,\eta)&=&0,
\end{eqnarray}
where
\begin{eqnarray}
Q(x)&\equiv&V(x)-E
\end{eqnarray}
is given by the ``potential'' $V$ and the ``energy'' $E$.
For the Landau-Zener model, we have
\begin{eqnarray}
Q(x,\eta)&=&\left(\Delta^2-i\eta^{-1}\frac{v}{2}\right)+\frac{1}{4}v^2t^2\nonumber\\
&=&\left(\Delta^2+\frac{1}{4}v^2t^2\right)+\left(\mp i\eta^{-1}\frac{v}{2}\right)\\
Q_0(x)&\equiv&\Delta^2+\frac{1}{4}v^2t^2\\
Q_{-1}(x)&\equiv&\mp i\eta^{-1}\frac{v}{2}.
\end{eqnarray}
Due to the formal structure of the
EWKB\cite{Enomoto:2020xlf,Virtual:2015HKT}, the Stokes lines are drawn
using only $Q_0$\footnote{See also Appendix \ref{app-srevEWKB} to find
the difference between the conventional WKB analysis and the EWKB.}.
Therefore, in the EWKB formulation, $\psi_1$ and $\psi_2$ 
have the same Stokes lines.
(A careful reader will understand that this statement does not mean that
solutions are identical.) 
Finally, we have
\begin{eqnarray}
V&=&-\frac{1}{4}v^2x^2\\
E&=&\Delta^2 
\end{eqnarray}
for the conventional quantum scattering problem with an inverted
quadratic potential.
See also Appendix \ref{app-reviewEWKB} and Ref.\cite{Enomoto:2020xlf}
for more details about the EWKB and the Stokes lines for cosmological particle production.

If one wants to consider (explicitly) the exact solution instead of the
Stokes lines of the EWKB, it will be convenient to consider 
$z=i\sqrt{v} e^{i\pi/4}t$
($z^2=-ivt^2$) to find\footnote{Here we temporarily set $\hbar=1$.}
\begin{eqnarray}
\left[\frac{d^2}{dt^2}+\left(n+\frac{1}{2}-\frac{1}{4}z^2\right)\right]\psi_1(z)&=&0\\
\left[\frac{d^2}{dt^2}+\left(n-\frac{1}{2}-\frac{1}{4}z^2\right)\right]\psi_2(z)&=&0.
\end{eqnarray}
Here we set 
\begin{eqnarray}
n&\equiv&i\frac{\Delta^2}{v}.
\end{eqnarray}
Since these equations are giving the standard form of the Weber
equation, their solutions are given by a couple of independent
combinations of $D_n(z), D_n(-z),D_{-n-1}(iz),
D_{-n-1}(-iz)$.
Using the asymptotic forms of the Weber function, one can easily get the
transfer matrix given by
\begin{eqnarray}
\left(
\begin{array}{c}
\psi_1^+\\
\psi_2^+
\end{array}
\right)&=&\left(
\begin{array}{cc}
e^{-\pi \kappa}& -\sqrt{1-e^{-2\pi\kappa}} \\
\sqrt{1-e^{-2\pi\kappa}} & e^{-\pi \kappa}
\end{array}
\right)
\left(
\begin{array}{c}
\psi_1^-\\
\psi_2^-
\end{array}
\right),\nonumber\\
\end{eqnarray}
where phase parameters are disregarded for simplicity.
$\pm$ signs of $\psi^\pm$ are for $t\rightarrow \pm \infty$.
We introduced $\kappa$, which is the imaginary part of $n$ and given by
\begin{eqnarray}
\kappa&\equiv&\frac{\Delta^2}{v}.
\end{eqnarray}
For the EWKB, this factor appears from the integral connecting the two turning
points of the MTP\cite{Enomoto:2020xlf}.
(Here, ``turning point'' denotes solutions of $Q_0=0$.)
For the cosmological particle production, $\kappa$ determines the number
density.
Note that the above transfer matrix is not defined for the ``adiabatic
states'', which represents the ``adiabatic energy'' 
\begin{eqnarray}
E_\pm&=&\pm\sqrt{\Delta^2+v^2t^2/4}.
\end{eqnarray}
Since these adiabatic states are diagonalizing the Hamiltonian and 
identified with the asymptotic WKB
solutions, the transition matrix for these (adiabatic) states is giving
Bogoliubov transformation of the cosmological particle production.
If one writes the transfer matrix for the ``adiabatic states''
$\Psi_{1,2}$ instead of the original states $\psi_{1,2}$, 
one will have
\begin{eqnarray}
\left(
\begin{array}{c}
\Psi_1^+\\
\Psi_2^+
\end{array}
\right)&=&\left(
\begin{array}{cc}
 \sqrt{1-e^{-2\pi\kappa}} &e^{-\pi \kappa}\\
e^{-\pi \kappa} &-\sqrt{1-e^{-2\pi\kappa}}
\end{array}
\right)
\left(
\begin{array}{c}
\Psi_1^-\\
\Psi_2^-
\end{array}
\right),\nonumber\\
\end{eqnarray}
where we have omitted the phase parameter.
Compare the transfer matrix with the one obtained for bosonic
preheating in Ref.\cite{Kofman:1997yn}.
For Dirac fermions, one can find the calculation based on the 
Landau-Zener model in Ref.\cite{Enomoto:2020xlf},
 which can be compared with the standard
calculation of Ref.\cite{Greene:1998nh, Peloso:2000hy}.
The off-diagonal elements of the transfer matrix are giving
$\beta_k^+$ of the Bogoliubov transformation\cite{Kofman:1997yn}
if $\alpha_k^-=1, \beta_k^-=0$ is considered for the initial condition.

Comparing the original equation of the Landau-Zener model and
 the decoupled equations,
one can see that $D_1\equiv - vt, D_2\equiv +vt$ in the (original) 
diagonal elements are transferred into the
 ``potential'' $-\frac{1}{4}v^2t^2$ in the decoupled
 equations\cite{Enomoto:2020xlf}.
In this paper, both approaches (the Landau-Zener model and the EWKB
 Stokes lines of the decoupled equations) are used to understand the cosmological particle
 production and the origin of the asymmetry.

\section{Asymmetry in cosmological particle production and the 
 decay process}
First, we introduce helicity-violating interaction (mass) for the
Majorana fermion and examine asymmetry production when the mass is time-dependent.
Because of the facility of the equations, our idea of asymmetric
particle production  will be examined first for the helicity asymmetry
of the Majorana fermions.
The result can be regarded as asymmetric decay of the $\theta(t)$ field
 (PNGB), where the asymmetry is determined by the sign of
$\dot{\theta}$.
Unlike the usual scenario of asymmetric decay, interference is not
playing important role in our scenario.

Although perturbative expansion before the non-perturbative analysis could be
very useful, perturbative expansion may drastically change the structure
of the EWKB Stokes lines of the model.
Therefore, perturbative expansion before the non-perturbative analysis
is sometimes very dangerous.
Some useful examples will be shown.

\subsection{Majorana fermion with time-dependent mass (Basic
  Calculation)}
\label{subsec-basic}
For the Majorana fermion, we consider $\Psi^t_R\equiv(\psi_R,\psi_R^\dagger)$
and write the Majorana mass term as
\begin{eqnarray}
{\cal L}_m&=&\bar{\Psi}_R
\left(
\begin{array}{cc}
   0   & m_R \\
   m_R^*   & 0 
\end{array}
\right)\Psi_R.
\end{eqnarray}
The Lagrangian density becomes
\begin{eqnarray}
\label{eq-majorana-Lag}
{\cal L}&=&
\bar{\psi}_Ri\bar{\sigma}^\mu\partial_\mu\psi_R -\frac{1}{2}
\left(m_R\psi_R^2+m_R^*\psi_R^{\dagger 2}\right),
\end{eqnarray}
which gives the equation of motion
\begin{eqnarray}
(i\bar{\sigma}^0\partial_t+i\bar{\sigma}^i\partial_i)\psi_R&=&
-m_R^*\psi_R^\dagger.
\end{eqnarray}
We consider the expansion
\begin{eqnarray}
(\psi_R)_\alpha&=&\int\frac{d^3k}{(2\pi)^3}e^{i\mathbf{k\cdot x}}\sum_{s=\pm}
(e^s_{\boldsymbol k})_\alpha \nonumber\\
&&\times \left[
u^s_k(t) a^s_{\boldsymbol k}
+v^{s*}_{k}(t)\cdot e^{-i\theta_{\boldsymbol k}}a_{-\boldsymbol k}^{s\dagger}\right],
\label{eq_expansion_MF}
\end{eqnarray}
where $e_{\mathbf{k}}^s$ denotes the helicity eigenstate and we have
\begin{equation}
 -k^i \bar{\sigma}^i e_{\mathbf{k}}^s = s|\mathbf{k}| \bar{\sigma}^0 e_{\mathbf{k}}^s
  \qquad (s=\pm) 
\end{equation}
and the orthogonalities
\begin{equation}
 e_{\boldsymbol k}^{s\dagger}\bar{\sigma}^0e_{\boldsymbol k}^{s'} = \delta^{ss'},
 \qquad e_{\boldsymbol k}^se_{- \boldsymbol k}^{s'}=se^{i\theta_{\boldsymbol k}}\delta^{ss'},
\end{equation}
and a phase by the momentum direction
\begin{equation}
 e^{i\theta_{\boldsymbol k}}\equiv \frac{k^1+ik^2}{\sqrt{(k^1)^2+(k^2)^2}}.
\end{equation}

From the above expansion and the equation of motion, one will find
\begin{eqnarray}
\label{eq-EOMofMajo}
(i\partial_t+s|\boldsymbol k|)u^s_{k}=s m_R^*v^{s}_{k},\nonumber\\
(i\partial_t+s|\boldsymbol k|)v^{s*}_{k}=-s m_R^* u^{s*}_{k}.
\end{eqnarray}
One can write these equations using a matrix as
\begin{eqnarray}
i\frac{d}{dt}\Psi&=&H\Psi,\nonumber\\
\left(
\begin{array}{cc}
H_{11} & H_{12}\\
H_{21} & H_{21}
\end{array}
\right)&=& 
\left(
\begin{array}{cc}
-s|\boldsymbol k| & s m_R^*(t)\\
s m_R(t) & s|\boldsymbol k|
\end{array}
\right),
\end{eqnarray}
where $\Psi^t\equiv (v^s_{k}, u^s_{k})$.

Note that this formalism introduces ``time-dependent off-diagonal
elements'' to the theory.
This is what we need for asymmetry production in this paper.

We are going to write the above equations in a versatile
form.
Introducing $D(t)=-s|{\boldsymbol k}|$ and $\Delta(t)=sm_R(t)$,
we have 
\begin{eqnarray}
\label{eq-simpleoriginalLZ}
i\hbar \frac{d}{dt}\left(
\begin{array}{c}
X\\
Y
\end{array}
\right)&=&\left(
\begin{array}{cc}
D(t) & \Delta(t)^*\\
\Delta(t) & -D(t)
\end{array}
\right)
\left(
\begin{array}{c}
X\\
Y
\end{array}
\right),
\end{eqnarray}
where we recovered $\hbar$ for later arguments.
Decoupling the equations, one will have the equations given by
\begin{eqnarray}
\ddot{X}-\frac{\dot{\Delta}^*}{\Delta^*}\dot{X}+
\left(-\frac{iD\dot{\Delta}^*}{\hbar\Delta^*}
+\frac{i\dot{D}}{\hbar}
+\frac{|\Delta|^2+D^2}{\hbar^2}
\right)X=0.\\
\ddot{Y}-\frac{\dot{\Delta}}{\Delta}\dot{Y}+
\left(\frac{iD\dot{\Delta}}{\hbar\Delta}
-\frac{i\dot{D}}{\hbar}
+\frac{|\Delta|^2+D^2}{\hbar^2}
\right)Y=0.
\end{eqnarray}
To obtain the standard form of the EWKB, one has to introduce a new
$\hat{P}$ and $\hat{Q}$ defined by
\begin{eqnarray}
\label{eq-normalEWKBtrans}
\hat{X}&=&\exp\left(-\frac{1}{2}\int^x
	       \frac{\dot{\Delta}^*}{\Delta^*}dx\right)X\nonumber\\
\hat{Y}&=&\exp\left(-\frac{1}{2}\int^x
	       \frac{\dot{\Delta}}{\Delta}dx\right)Y
\end{eqnarray}

Alternatively, one can introduce $\hat{P}$(and $\hat{Q}$) to the
original equation (before the decoupling) to remove the time-dependence
of the off-diagonal elements in (\ref{eq-simpleoriginalLZ}).
In that case, one will have
\begin{eqnarray}
\Psi&=&U \hat{\Psi},\\
\hat{H}&=&U^{-1} H U-i U^{-1}\dot{U},
\end{eqnarray}
where $U$ defines the transformation given in Eq.(\ref{eq-normalEWKBtrans}).
Of course, after decoupling the equations, one will have identical results.

For the decoupled equations, the equations can be written in the standard EWKB form
\begin{eqnarray}
\label{Eq_2ndorder-Majorana}
\ddot{\hat{X}}&+&\left(\frac{-iD\dot{\Delta}^*}{\hbar \Delta^*}
+\frac{i\dot{D}}{\hbar}+\frac{|\Delta|^2+D^2}{\hbar^2}\right.\nonumber\\
&+&\left.\frac{\ddot{\Delta}^*}{2\Delta^*}-\frac{3(\dot{\Delta}^*)^2}{4(\Delta^*)^2}\right)\hat{X}=0\\
\ddot{\hat{Y}}&+&\left(\frac{iD\dot{\Delta}}{\hbar \Delta}
-\frac{i\dot{D}}{\hbar}+\frac{|\Delta|^2+D^2}{\hbar^2}\right.\nonumber\\
&+&\left.\frac{\ddot{\Delta}}{2\Delta}-\frac{3(\dot{\Delta})^2}{4(\Delta)^2}\right)\hat{Y}=0.
\end{eqnarray}

Seeing the $\hbar$-dependence, the EWKB Stokes lines of the above equation
coincides with the trivial equation
\begin{eqnarray}
\label{eq-triv}
\ddot{\hat{X}}&+&\frac{|\Delta|^2+D^2}{\hbar^2}\hat{X}=0,
\end{eqnarray}
which cannot generate asymmetry.
Mathematically, this result is true if no extra $\hbar$ is appearing from $\dot{\Delta}^*$.
We are going to examine this
model further to understand the source of the asymmetry.

\subsection{Majorana fermion with time-dependent mass (constant rotation)}
\label{subsec-simplept}
We start with the typical example of quantum mechanics.
In cosmology, trapping of an oscillating field\cite{Kofman:2004yc,
Enomoto:2013mla} or the Affleck-Dine baryogenesis\cite{Affleck:1984fy,
Matsuda:2002jv} can generate similar rotation, which can be seen in a
local domain\cite{Lee:1991ax,Matsuda:2002jx}.

Here we consider the equation given by\footnote{Just for simplicity, we
temporarily set $\hbar=1$.}
\begin{eqnarray}
i\frac{d}{dt}\Psi&=&(H^{(0)}+H^{(1)})\Psi,\nonumber\\
H^{(0)}&=& 
\left(
\begin{array}{cc}
D& 0\\
0& -D\\
\end{array}
\right)\\
H^{(1)}&=& 
\left(
\begin{array}{cc}
0 & \Delta^*(t)\\
\Delta_1(t)& 0\\
\end{array}
\right),
\end{eqnarray}
where $D= \omega_0$.
Then we have the solution given by
\begin{eqnarray}
\Psi(t)&=&c_1(t) e^{-i\omega_0 t} \left(
\begin{array}{c}
1\\
0
\end{array}
\right)
+c_2(t) e^{i\omega_0 t} \left(
\begin{array}{c}
0\\
1
\end{array}
\right).
\end{eqnarray}
To find the time-dependent coefficients $c_{1}(t)$ and
$c_2(t)$, we substitute $\Psi(t)$ to find
\begin{eqnarray}
i\frac{d c_1}{dt}&=& \Delta^* e^{2i\omega_0 t}c_2(t)\\
i\frac{d c_2}{dt}&=& \Delta e^{-2i\omega_0 t}c_1(t).
\end{eqnarray}
It is very difficult to solve this equation exactly for general $\Delta(t)$, 
but one can use a numerical calculation to understand the transition.
For $\Delta(t)\equiv Ae^{i\omega t}= Ae^{2i\omega_0 t}$, one can easily find the exact
solution, which gives $c_{1,2}(t)\sim \sin (At+\theta_0)$.
Here $A$ ($=|\Delta|$) determines the rapidity of the transition and the
maximum transition is possible for any $A$, although it takes a long time for small $A$.
Away from the resonance frequency at $\omega=2\omega_0$, the transition amplitude
decreases.

Here, what is important for our discussion about asymmetry is
inverse rotation.
If the off-diagonal element is replaced by 
$\Delta(t)= Ae^{-2i\omega_0t}$, the equations become
\begin{eqnarray}
i\frac{d c_1}{dt}&=& A e^{4i\omega_0 t}c_2(t)\\
i\frac{d c_2}{dt}&=& A e^{-4i\omega_0 t}c_1(t),
\end{eqnarray}
which ruins the resonance.

\begin{figure}[t]
\centering
\includegraphics[width=0.9\columnwidth]{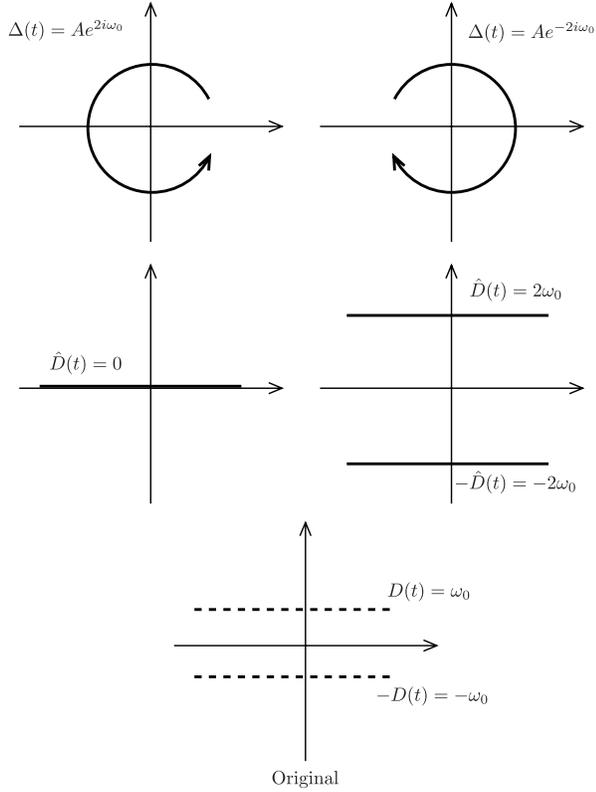}
 \caption{Top: rotation of the off-diagonal elements.
Middle: New states of $\hat{H}$.
Bottom: The original states of $H$.
In this case, the resonance is possible only for the anti-clockwise rotation.}
\label{fig-rot-simplest}
\end{figure}

The situation becomes very clear if one introduces the transformation of
Eq.(\ref{eq-normalEWKBtrans}).
For $\Delta(t)= Ae^{2i\omega_0t}$, the two states are shifted together
to make a pair of degenerated states ($\hat{D}=0$) in $\hat{H}$.
If this transition corresponds to the Bogoliubov transformation,
particle production is possible in this case.
On the other hand, for the inverse rotation $\Delta(t)=
Ae^{-2i\omega_0t}$, two states are ``shifted away'' to give
$\hat{D}=2\omega_0$.
Then the particle production is suppressed.
Although the diagonal elements of $\hat{H}$ coincide, the ``adiabatic''
energy splits because of the remaining off-diagonal
elements (i.e, the radial part $|\Delta(t)|$), which affects the rapidity of the
transition.
The situation is shown in Fig.\ref{fig-rot-simplest}.

Although this model is very useful for understanding the origin of the
asymmetry, there is a problem in defining the number density.
In this model, there is no non-perturbative transition (i.e, the Stokes phenomena)
between the ``adiabatic states''.
On the other hand, the adiabatic states cannot keep diagonalizing the
Hamiltonian, which causes oscillation of the number density.
To avoid such problem, one has to define the creation and the
annihilation operators for the asymptotic states, but in this model the
asymptotic states are not defined.\footnote{
In reality, the period of the periodic function $n_k$ depends on $k$.
Therefore, the total number density after $k$-intergation
is not simple.
Since the particles could decay during the process, it could be possible
to claim that asymmetric particle production is possible in this model.}.
To introduce a non-adiabatic transition, we have to extend the model.

\subsection{Majorana fermion with linear time-dependence ($\Delta=s g
(\epsilon+i v t), D=-s|\boldsymbol k|$)} 
\label{subsec-lin}
For the cosmological preheating scenario, particle production near the
enhanced symmetry point(ESP) with 
$\phi(t)=\epsilon+ivt$ is very important.
Using the generation of the asymmetry, we also explain the crucial
discrepancy between the conventional WKB expansion and the EWKB.

We start with the decoupled equation
\begin{eqnarray}
\ddot{\hat{X}}&+&\left[\frac{g^2 \left(\epsilon^2+v^2t^2 \right)+
		  |\boldsymbol 
		  k|^2}{\hbar^2}+\frac{s|\boldsymbol k| v}{\hbar
		  (\epsilon-ivt)}\right.\nonumber\\ 
&&\left.
+\frac{3v^2}{4(\epsilon-ivt)^2}\right]\hat{X}=0.
\end{eqnarray}
Naively for the EWKB, this equation gives 
\begin{eqnarray}
Q_0(t)&=&-g^2 \left(\epsilon^2+v^2t^2 \right)- |\boldsymbol k|^2
\end{eqnarray}
and the asymmetry seems to disappear from the non-perturbative
calculation.
This result agrees with our numerical calculation.
However, seeing the trajectory, one will find that this model has a
rotational motion around the origin (i.e, the ESP).
Is it true that the rotational asymmetry considered in
Sec.\ref{subsec-simplept} disappears in this model?
If the asymmetry disappears, what is the crucial
condition?\footnote{In Ref.\cite{Dolgov:1989us}, by considering
perturbation, the asymmetry is related to the interference between terms.
We are arguing this topic from another viewpoint.}

\begin{figure}[t]
\centering
\includegraphics[width=0.9\columnwidth]{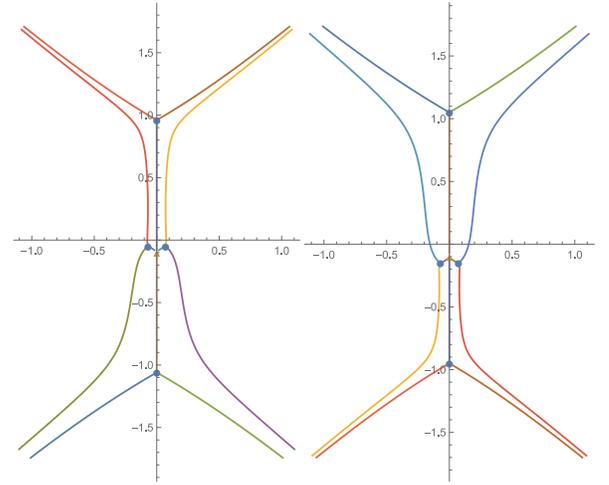}
 \caption{Stokes lines for $\epsilon=0.1, v=1, k=1, \hbar=0.1,
 g=1, s=\pm1$ are
 given for $Q_0+Q_1$, not for $Q_0$.
One can see four turning points (circle) and a double pole (triangle)
 near the origin, which are asymmetric, 
but they disappear in the EWKB.}
\label{fig_stokesdoublepoll}
\end{figure}

To understand more about the situation, we show the naive Stokes
lines\footnote{The stokes lines are called ``naive'', 
since careful people will not use these stokes lines for their
calculation, even if they are 
considering the conventional WKB expansion.
The problems of $O(\hbar)$ terms and their poles are widely
known\cite{Berry:1972na} for the conventional WKB expansion.}
in Fig.\ref{fig_stokesdoublepoll}, which have a double pole at
$t=-i\epsilon /v$ and four turning points.\footnote{Note that
the Stokes lines in Fig.\ref{fig_stokesdoublepoll} are 
obtained from $Q$ itself, not from $Q_0$.
Therefore the stokes lines in Fig.\ref{fig_stokesdoublepoll} 
are not representing the true stokes lines.}
Seeing the Stokes lines, one can understand the situation.
The EWKB stokes lines appear after gluing the double pole and two
turning points together at the origin. 
In this limit (i.e, $\hbar \rightarrow 0$), the stokes lines give the
simple ``scattering problem with an inverted quadratic
potential''\cite{Enomoto:2020xlf}.

Alternatively, using the numerical calculation explained in
Appendix \ref{sec_distribution_MF}, 
we find that the net number $n_k^+-n_k^-$ cannot evolve in the straight
motion $\ddot{m}_R=0$
as long as the initial state begins with the zero particle state.
In reality, however, the backreaction is
important since the particle production is significant in this
model, and its backreaction can bend the trajectory to give
$\ddot{m}_R\ne 0$.\footnote{In the most
significant case, the oscillating field can be ``trapped'' near the
ESP\cite{Kofman:2004yc}, which may happen also for higher-dimensional interaction suppressed by the Planck
scale\cite{Enomoto:2013mla}}.

Therefore, without considering the backreaction, this model cannot
generate asymmetry, but the particle production always introduces the
backreaction, which is required for the asymmetry.
Therefore, in reality, the asymmetry production cannot be neglected in this model.

Below, we mainly consider only the rotational motion of the
 off-diagonal elements, which depends only on the phase $\theta(t)$.
In such models, the Landau-Zener transition is very useful.
We are using the EWKB to support the analysis.

\subsection{Majorana fermion with time-dependent mass (perturbative expansion)}
\label{sec-pert}
We consider perturbative expansion of the model considered in
Sec.\ref{subsec-simplept}, which is given by
\begin{eqnarray}
\Delta&=&A e^{i\theta}\simeq A (1+i\theta+...),
\end{eqnarray}
where $\theta(t)=\omega t$ is the simplest example.
The expansion\footnote{The
EWKB is an expansion, but it takes the Borel sum to get the 
non-perturbative result.}
is valid only for $\theta\ll 1$.
Alternatively, one can choose $\theta(t)=A \cos \omega t$ with $A\ll 1$.

Reflecting on the calculation above, 
 the above perturbative expansion might have changed the mechanism of
 the non-perturbative process.
The significant discrepancy appears when $\theta = \omega t$.
Indeed, the original theory does not allow non-adiabatic transition,
while in the perturbed theory the transition is solved as
 the quantum scattering by a potential (after decoupling the equations).
Therefore, in this example, one has to conclude that the essential
mechanism of the transition has been changed by
taking the perturbative expansion.
In the light of the EWKB, this is because the structure of the
Stokes lines, which 
determines the transfer matrix, is changed by the perturbation.
Therefore, to avoid such discrepancy, one has to consider the expansion
that does not change the essential property of the Stokes
lines.
We show some typical examples in appendix\ref{app-reviewEWKB}.

If the particle production is the Landau-Zener type, one has to check
first the global structure of the Stokes lines of the EWKB, and the
local (linear) expansion has to be taken around the points where the
Stokes lines cross the real-time axis. 
Indeed, the original Landau-Zener model considers local expansion around the
state-crossing point, where the global structure of the Stokes lines has
the separable form of the MTP(Merged pair of simple Turning Points).
In this paper, we often consider similar expansion.
Note however that the particle production is not always explained by the
Landau-Zener type transition.

We sometimes compare the Landau-Zener type calculation with the EWKB
Stokes lines.
In our analysis, the Landau-Zener type transition is very
useful for understanding the asymmetry.

\subsection{Majorana fermion with time-dependent mass
\label{sec-majcos}
(The EWKB for $\Delta(t)=m_0 e^{i\theta}$, $\theta(t)\equiv A\cos(\omega_a t/\hbar)$)}
\label{sec-oscphase}
In this section, we consider
\begin{eqnarray}
\theta(t)&\equiv& A\cos(\omega_a t/\hbar)\nonumber\\
\Delta(t)&=&m_0 e^{-i\theta}
\end{eqnarray}
and draw the Stokes lines of Eq.(\ref{Eq_2ndorder-Majorana}).
Note that
\begin{eqnarray}
\frac{\dot{\Delta}^*}{\Delta^*}&=&i\dot{\theta}\\
&=&\frac{i\omega_a}{\hbar}\sin\left(\frac{\omega_a t}{\hbar}\right)\nonumber\\
&\sim&O(\hbar^{-1})
\end{eqnarray}
is important for the EWKB calculation.\footnote{In the EWKB, analytic
evaluation of the integral is usually very difficult, despite the
transparency of the qualitative analysis based on the Stokes lines.
In this paper, we are using the EWKB for qualitative analysis.}

What is important for the asymmetry is the term proportional to
$s$.
Terms proportional to $\dot{D}$ will disappear since $D$ is constant.
Therefore, the s-dependence comes from the factor
\begin{eqnarray}
-iD\frac{\dot{\Delta}^*}{\hbar\Delta^*}&=&
-\frac{i(-s|\boldsymbol k|)}{\hbar}\frac{i\omega_a}{\hbar}\sin\left(\frac{\omega_a t}{\hbar}\right)\nonumber\\
&=&
-\frac{s|\boldsymbol k|\omega_a}{\hbar^2}
\sin\left(\frac{\omega_a t}{\hbar}\right).
\end{eqnarray}
This term is suppressed in the non-relativistic limit ($|\boldsymbol
k|\rightarrow 0$).
Therefore, the asymmetry is not significant in the non-relativistic limit.
This result may not be consistent with the intuition, since 
in the non-relativistic limit, violation of the helicity could be
significant.
We will explain the reason later in this section.

To understand more about the origin of the asymmetry, we have calculated the Stokes
lines for typical $Q_0$, which is shown in Fig.\ref{fig_exactpm}
together with their ``potentials''.
We have excluded the imaginary part for simplicity.
See Appendix \ref{app-reviewEWKB} for more details.
\begin{figure}[t]
\centering
\includegraphics[width=0.9\columnwidth]{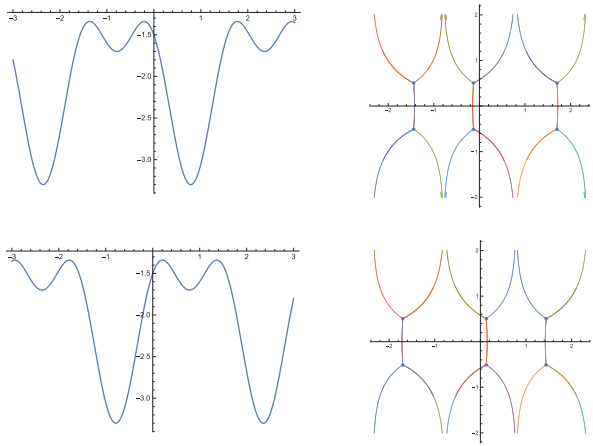}
 \caption{The Stokes lines and the potential are shown for 
$Q(t)= -1.5 - s 0.8 \sin(2t) - \sin^2(2t)$ and
$V(t)=- s 0.8 \sin(2t) - \sin^2(2t)$.
The upper is for $s=+1$ and the lower is for $s=-1$.}
\label{fig_exactpm}
\end{figure}
Seeing Fig.\ref{fig_exactpm} and the solutions of $Q_0=0$, one can
understand that the basic structures of the MTPs\footnote{In terms of the
EWKB, each MTP gives the transfer matrix similar to the 
scattering by an independent inverted quadratic potential, which is
calculable\cite{Enomoto:2020xlf,Sueishi:2020rug, Taya:2020dco, Sueishi:2021xti}.
In this sense, each MTP is separable. See also Appendix
\ref{app-reviewEWKB} for more details.} are completely 
the same but their positions are shifted by changing the sign of $s$.
Therefore, for an eternal oscillation, the production becomes
symmetric on average, but for a damped oscillation, the asymmetry
appears because  
the particle production is not simultaneous for different $s=\pm1$.
The time-dependent amplitude of the oscillation
 generates different number densities at each MTP.
The asymmetry is therefore determined by the time-dependence of the
amplitude.
Particle production is exclusive at each MTP\footnote{
When $s=1$ is generated, the other ($s=-1$) is not generated.
Simultaneous production is possible only when $k=0$, where the asymmetry
vanishes. 
Of course, particle production is not instant in reality.}

\begin{figure}[t]
\centering
\includegraphics[width=0.9\columnwidth]{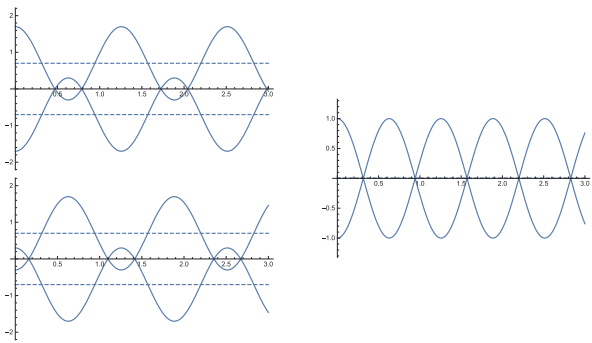}
 \caption{After using Eq.(\ref{eq-normalEWKBtrans}), 
the time-dependence appears only in the diagonal elements.
Left: As far as $|\boldsymbol k|\ne 0$, one can discriminate $s=1$ and
 $s=-1$.
Right: For $|\boldsymbol k|= 0$, particle production
 cannot discriminate the helicity.}
\label{fig_cosk}
\end{figure}

One thing that has to be clarified is the vanishing asymmetry in the
limit of $|{\boldsymbol k}|\rightarrow 0$.
This phenomenon can be understood easily by considering the original
equation (Landau-Zener type).
In Fig.\ref{fig_cosk}, we show the motion of the two states for $\hat{H}$.
In the left, $s=\pm 1$ states are distinguishable when $|{\boldsymbol
k}|\ne 0$, while in the right, $s=\pm 1$ states are not 
distinguishable when $|{\boldsymbol k}|= 0$.
Simultaneous particle production is possible only when $k=0$, which
corresponds to vanishing asymmetry.

One can compare the above result with the earlier calculation given in
Ref.\cite{Dolgov:1996qq, Enomoto:2018yeu}.
In Ref.\cite{Dolgov:1996qq}, perturbative expansion
has been used to expand $e^{i\theta}$ to calculate non-perturbative
particle production, where the asymmetry appears from the interference
between terms.
In our analysis, the asymmetry appears because the particle
production is exclusive.\footnote{Note however that
the asymmetry depends on $k$.
One can see from Fig.\ref{fig_cosk} that states are symmetric for $k\simeq 0$.
Therefore, total number density after $k$ integration may not be utterly
exclusive due to the symmetric particle production near $k\sim 0$.}

In Ref.\cite{Enomoto:2018yeu}, it has been shown for a single Dirac fermion
that crucial cancellation disturbs asymmetry production.
The reason is very simple.
For a single Dirac fermion, the complex mass has to be introduced as
\begin{eqnarray}
m_D \overline{\psi_L}\psi_R +m_D^* \overline{\psi_R}\psi_L,
\end{eqnarray}
which flips the rotational direction by the exchange $L\leftrightarrow R$.
More precisely, since the exchange $L\leftrightarrow R$ inverses the
rotational motion of the complex Dirac mass, ``matter production''
of the left-handed fermion occurs simultaneously with ``antimatter production'' of
the right-handed fermion, and vice versa.
To avoid this cancellation, one has to introduce more than two Dirac
fermions, whose off-diagonal elements (interaction) is given by 
\begin{eqnarray}
\left[m_{\Delta} (\overline{\psi_{Lj}}\psi_{Ri}
 +\overline{\psi_{Rj}}\psi_{Li})+h.c.\right].
\end{eqnarray}
or
\begin{eqnarray}
\left[m_{\Delta} \overline{\psi_{Lj}}\psi_{Ri}
 +h.c.\right].
\end{eqnarray}

We will go back to this topic in Sec.\ref{sec-concex}.

\subsection{Majorana fermion with time-dependent mass
($\theta(t)=A/(1+e^{-at/\hbar})$)}
\label{subsec-onetime}
\begin{figure}[t]
\centering
\includegraphics[width=0.9\columnwidth]{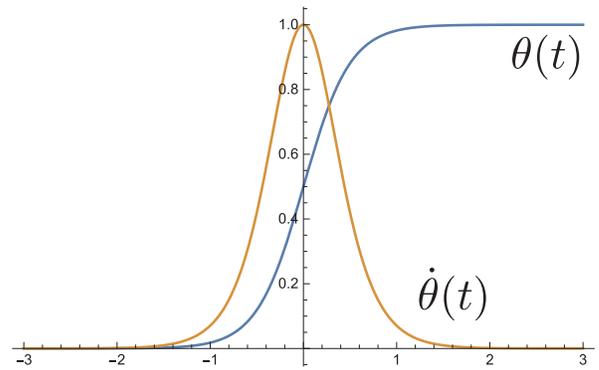}
 \caption{$\theta(t)$ and $\dot{\theta}(t)$ for $\theta(t)=A/(1+e^{-at/\hbar})$}
\label{fig_trans-easy1}
\end{figure}

In this section, we consider the simple rotational motion of the phase.
As we will see later in this section, the model clearly describes the origin of
the asymmetry in particle production.
We consider
\begin{eqnarray}
\theta(t)=\frac{A}{1+e^{-at/\hbar}},
\end{eqnarray}
which gives transition of the phase from $\theta_i=0$ to $\theta_e=A$
around $t=0$.
We showed the motion of $\theta(t)$ and $\dot{\theta}(t)$
in Fig.\ref{fig_trans-easy1}.
We take 
\begin{eqnarray}
m_R(t)&=&g\varphi (t)\equiv g\varphi_0 e^{i\theta(t)}\nonumber\\
\theta(t)&\equiv&\frac{A}{1+e^{-at/\hbar}},
\end{eqnarray}
where $\varphi_0$ is real.
Again, one can remove the time-dependence of the off-diagonal
elements using the translation\footnote{See Eq.(\ref{eq-normalEWKBtrans}).
This transformation was originally introduced to get the standard form of the EWKB
formulation from the decoupled equations.
At the same time, it removes the time dependence of the off-diagonal
elements.}

\begin{eqnarray}
\hat{\Psi}&\equiv& U_T^{-1} \Psi,\nonumber\\
U_T&\equiv& 
\left(
\begin{array}{cc}
e^{i\theta(t)/2} & 0\\
0 & e^{-i\theta(t)/2}
\end{array}
\right).
\end{eqnarray}
The Hamiltonian after the transformation becomes
\begin{eqnarray}
\hat{H}&\equiv& U_T^{-1}H U_T -iU_T^{-1}\dot{U}_T\nonumber\\
&=& 
\left(
\begin{array}{cc}
-s|\boldsymbol k|-\frac{1}{2} \hbar\dot{\theta} & s g\varphi_0\\
s g^* \varphi_0  & s|\boldsymbol k|+\frac{1}{2} \hbar\dot{\theta}
\end{array}
\right).
\end{eqnarray}
Note that the new terms in the diagonal elements do not have ``$s$'' in
their coefficients.

\begin{figure}[t]
\centering
\includegraphics[width=0.9\columnwidth]{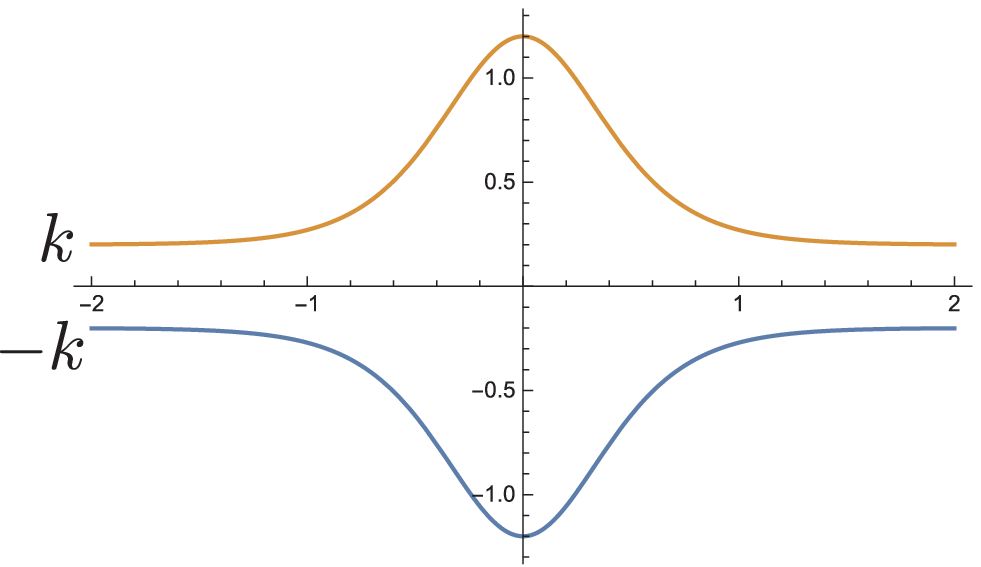}
\includegraphics[width=0.9\columnwidth]{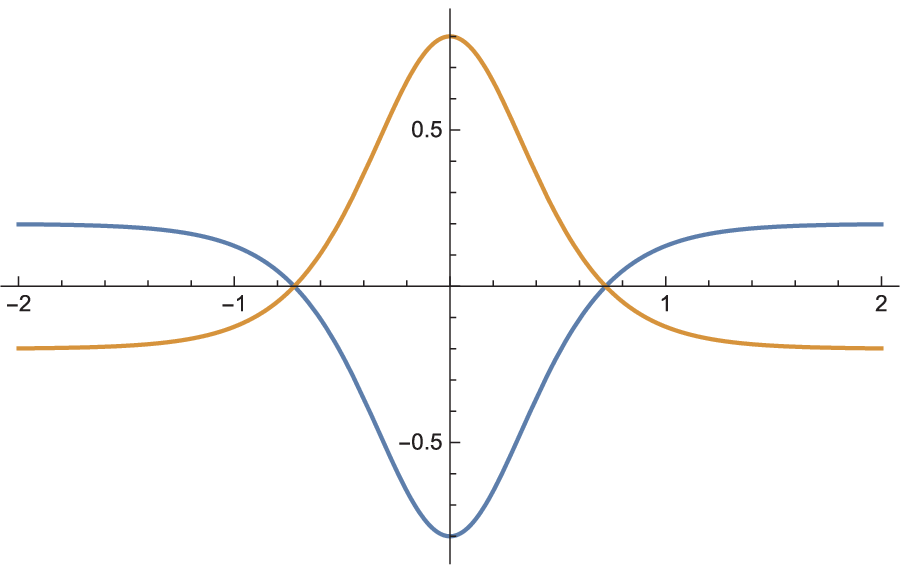}
 \caption{Upper: Time-dependence of the diagonal elements of
 $\hat{H}$ for $s=+1$ and $k\ne 0$, in which one can see no crossing.
Lower: Time-dependence of the diagonal elements of
 $\hat{H}$ for $s=-1$ and $k\ne 0$, in which each state crossing causes Landau-Zener
 transition.}
\label{fig_trans-easy}
\end{figure}

The new Hamiltonian is quite useful for our discussion.
One can see in Fig.\ref{fig_trans-easy} that the new term ($\propto
\dot{\theta}$) makes a bump around $t=0$ and the intersection (in the
sense of the original Landau-Zener model) appears only for $s=-1$,
because of the signs in front of $\dot{\theta}$.
This means that for a certain range of $k$, only $s=-1$
particles can experience the Landau-Zener type transition.
The particle production seems exclusive in this case, at least for the
Landau-Zener type particle production.\footnote{In appendix
\ref{app-reviewEWKB}, we argue the possibility of particle production
without crossing (not the Landau-Zener type).}
In this model, one can see state crossing of $s=-1$ for 
$k<k_*\equiv \frac{aA}{8}$.
Since $A$ defines the amplitude of $\theta(t)$, we found $n\propto
k_*^3\propto A^3$ in this model, which supports the claim
given in Ref.\cite{Dolgov:1996qq}.\footnote{The required condition for the
state crossing is calculated from the maximum of $\dot{\theta}$ (i.e,
the height of the bumps).
For $|k|>\hbar \mathrm{Max}[\dot{\theta}]/2 =|\hbar \dot{\theta}(0)|/2=a A/8$,
states are always apart.}

Considering the original Landau-Zener model, 
the probability of the transition (i.e, the magnitude of the particle
production) is determined by the velocity at each crossing point
($t_{1,2}$), which is
\begin{eqnarray}
\frac{|g\varphi_0|^2}{v(t_i)}&<&1\\
v(t_i)&\equiv& \dot{\hat{D}}(t_i)\equiv\hbar\frac{\ddot{\theta}(t_i)}{2}.
\end{eqnarray}
Note that in this formulation, $k$-dependence is appearing in $t_i$.

\begin{figure}[t]
\centering
\includegraphics[width=0.9\columnwidth]{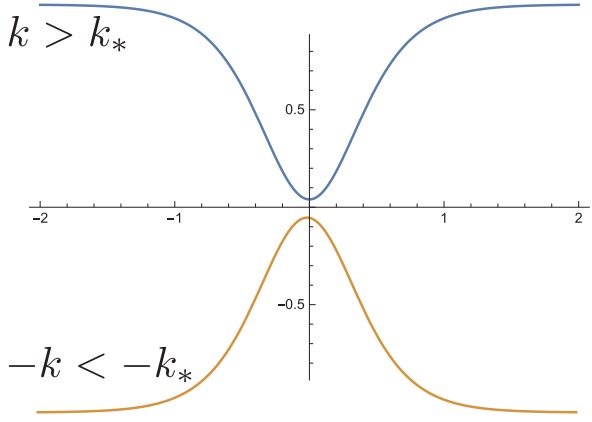}
 \caption{The upper and the lower states start to fall apart for larger
 $|k|$, and the crossing disappears near $k\sim k_*$.
At this moment $\ddot{\theta}$ disappears and higher calculation becomes
 important\cite{Enomoto:2020xlf}.}
\label{fig_trans-noteasy}
\end{figure}

In this argument, the radius of the Fermi sphere is important for
estimating the number density.
For large $k$, the top of the two bumps meet at
\begin{eqnarray}
k_*&\equiv&\frac{aA}{8},
\end{eqnarray}
where the velocity of the state becomes $\ddot{\theta}\simeq 0$.
See the Fig.\ref{fig_trans-noteasy}.
Here the conventional Landau-Zener model, which uses the linear expansion
at the state-crossing point, does not give the correct answer.
The transition of this kind has already been calculated in
Ref.\cite{Enomoto:2020xlf}, in which the scattering problem with an
inverted quartic potential has been considered.
Near the top of the bumps, we considered in Ref.\cite{Enomoto:2020xlf}
the equation given by
\begin{eqnarray}
i\hbar\frac{d}{dt}\left(
\begin{array}{c}
\psi_1\\
\psi_2
\end{array}
\right)&=&\left(
\begin{array}{cc}
-(a t^2+\epsilon)& \Delta \\
 \Delta& at^2+\epsilon 
\end{array}
\right)
\left(
\begin{array}{c}
\psi_1\\
\psi_2
\end{array}
\right),
\end{eqnarray}
which gives after decoupling the following equation
\begin{eqnarray}
\left[\hbar^2 \frac{d^2}{dt^2}+\left(\Delta^2-i(2at)\hbar\right)+\left(at^2+\epsilon\right)^2\right]\psi_1&=&0.
\end{eqnarray}
In terms of the EWKB, $Q_0$ of the equation is 
\begin{eqnarray}
\label{eq-fermion1}
Q_0&=&-\Delta^2-\left(at^2+\epsilon\right)^2,
\end{eqnarray}
where the ``potential'' is 
\begin{eqnarray}
V(t)&=&-a^2t^4-2a\epsilon t^2<0
\end{eqnarray}
and the ``energy'' is $E=\Delta^2+\epsilon^2>0$.
We thus find
\begin{eqnarray}
Q_0(t)&=&-\Delta^2-\epsilon^2- a^2 t^4-2a\epsilon t^2.
\end{eqnarray}

For the model considered in this section, we have
\begin{eqnarray}
\dot{\theta}(t)&\simeq&
2k_*+\frac{1}{2}\dddot{\theta}(0)t^2+...\\
\dddot{\theta}(0)&\equiv& -4a_2,
\end{eqnarray}
which gives the Hamiltonian for $s=-1$
\begin{eqnarray}
\hat{H} &=& 
\left(
\begin{array}{cc}
\Delta k +a_2t^2 & - g\varphi_0\\
- g^* \varphi_0  & -\Delta k -a_2t^2
\end{array}
\right),
\end{eqnarray}
where we introduced $\Delta k \equiv |k|-k_*$ .
There is no state crossing for $|k|>k_*$($\Delta k>0$), but to solve the
scattering problem of the quartic potential, one can see that 
the transition can be significant yet\cite{Enomoto:2020xlf}.
Transforming the above equation into $Q(x)=-\kappa_4 -x^4$, one can easily
find that $\kappa_4$ in the equation gives the bound $\kappa_4\lesssim 1$, which
determines the radius of the Fermi sphere.

Although rather crude, one can see that when the tops of the bumps coincide 
at $\Delta k=0$, the Fermi sphere can be larger than
$k_*$ if $a \ge (g^2\varphi_0^2+\Delta k)^{1/2} A^{1/3}$ is satisfied.
Here, $A$ represents the variation (amplitude) of $\theta(t)$ and $a$
determines the speed of the transition (or the width of the domain wall).
The above condition is showing that particle production is more efficient
when the variation of $\theta$ is larger and the speed of the transition is
higher.
This result coincides with intuition.

For $A\sim O(1)$, $a\sim g\varphi_0$ gives the Fermi sphere
$k_F=k_*\simeq g\varphi_0 /8$.
Variation from this point will change the Fermi sphere from the naive
estimation $k_F\sim k_*$.

\begin{figure}[t]
\centering
\includegraphics[width=1.0\columnwidth]{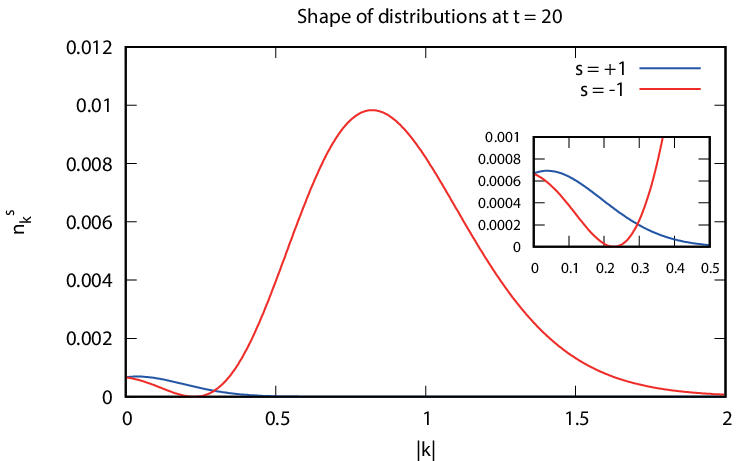}
\includegraphics[width=1.0\columnwidth]{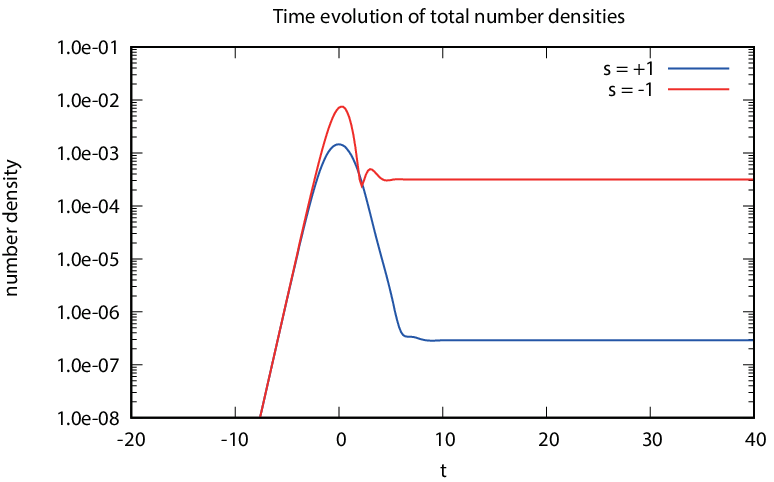}
 \caption{The results of numerical calculation with
 the parameters $|m_R|=a=1.0, A=\pi/3.$
 Upper: shapes of the distributions for each helicity state after the production ($t=20$).
 Lower: time evolution of the number densities (log scale).}
\label{fig_majorana_numerical}
\end{figure}

We show our numerical results\footnote{
In our numerical calculation,
the wave functions $u_k^s$ and $v_k^s$ appeared in (\ref{eq_expansion_MF})
are solved numerically, and they are interpreted into the distribution function.
In Appendix \ref{sec_distribution_MF} we show how to calculate the distribution function
by the wave functions.} in Fig.\ref{fig_majorana_numerical},
which depict the shapes of the distributions (upper panel) and the time evolution
of the number density for each state (lower panel).
One can see clearly that $s=+1$ production is suppressed
compared with $s=-1$. (Note that in the lower panel the number densities are given for the
log scale.)

\begin{figure}[t]
\centering
\includegraphics[width=0.9\columnwidth]{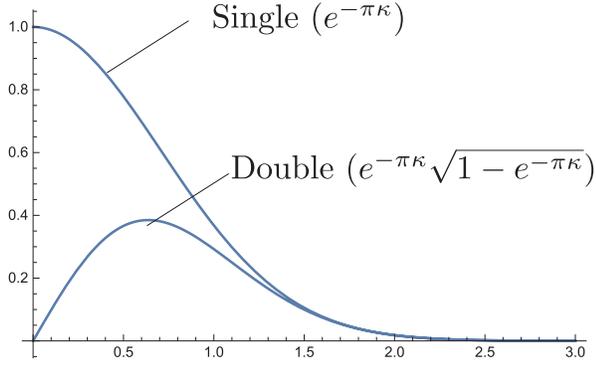}
 \caption{Typical distribution functions for $\kappa(\Delta)$ are shown for
 the single and the double MTP models.}
\label{fig_double-cross}
\end{figure}

In this model, the MTP structure always appears twice during
particle production.
When one calculates the Fermi sphere of particle production, this
double MTP structure causes a less trivial modification of the distribution,
as far as the particles do not decay between the two MTPs.
The typical distributions are shown in Fig.\ref{fig_double-cross} for
$\kappa(\Delta)\propto \Delta^2$.
Let us explain why this distribution is very important for our scenario.
Since the off-diagonal elements are representing the symmetry-violating
interaction, asymmetry in the $g\varphi_0\rightarrow 0$ limit is very
important.
For the Majorana fermion, since the off-diagonal elements
are representing the Majorana mass, small $|\Delta|$ seems to be making
the particle production easier.
This speculation is true for a single MTP scenario, but for our double
MTP scenario, one has to consider Fig.\ref{fig_double-cross} for the
distribution.
In the $g\varphi_0\rightarrow 0$ limit, the particle production vanishes
as one can see from Fig.\ref{fig_double-cross}.
Therefore, a tiny interaction suppresses particle production (and asymmetry).

From the above results, one can see that static motion
(always $\ddot{\theta}=0$) gives no state crossing in this model.

Although very crude, one can estimate the particle production when
cosmological domain walls decay.
For the typical cosmological domain wall of tension $\sigma \sim \Lambda^3$ and
 width $\delta_w\sim \Lambda^{-1}$ decaying at the density $\rho\sim \sigma^2/M_p^2$, the
 number density of the produced particle can be estimated as
 $k_F\sim 0.1 \Lambda$ for fast-moving wall.
However, since efficient particle production causes friction to the domain wall
motion, the maximum number density should be estimated as $n\lesssim
\rho/\Lambda\sim \Lambda^5/M_p^2$ when the temperature of the Universe
is $T_D\sim \sqrt{\sigma/M_p}$.
We conclude that asymmetric particle production can be
efficient for typical cosmological domain walls.

\section{Dirac fermions}
\label{sec-concex}
As we have mentioned previously, asymmetry will be canceled for a single Dirac fermion.
Let us see more details of the cancellation mechanism.
For a single Dirac fermion, one may find complex mass terms as
\begin{eqnarray}
\left[m_{\Delta} \overline{\psi_{L}}\psi_{R}
 +m_{\Delta}^* \overline{\psi_{R}}\psi_{L}\right]
\end{eqnarray}
Following the strategies we have considered in this paper, one can
explicitly calculate particle production for four species; 
left(right)-handed matter(antimatter).
Our observation is very simple.
Since the $L\leftrightarrow R$ exchange flips the rotational direction, 
one can immediately find that $L$ and $R$ productions are exclusive.
Since the matter-antimatter is also exclusive, L-matter production and
R-antimatter production are simultaneous.
Regarding this particle production as the decay process, 
one can say that depending on the sign of $\dot\theta$,
 the field $\theta$ ``selectively'' decays into L-matter and
 R-antimatter
(or L-antimatter and R-matter for the opposite sign).
Therefore, for a single Dirac fermion, direct asymmetry production is
impossible in total.

Although a single Dirac fermion is excluded for the (direct)
baryogenesis, there is hope in multiple Dirac fermions.
To show an example, let us consider a fictitious ``lepton'' $\psi^l$ and a
``quark'' $\psi^q$ for the calculation and suppose that each
fermion is massless (i.e, we are considering a naive relativistic limit)
but they have the Yukawa interaction 
\begin{eqnarray}
\label{eq-QLint}
&&g\phi \overline{\psi^l}\psi^q + h.c\nonumber\\
&=&g\phi(\overline{\psi^l_R}\psi^q_L+\overline{\psi^l_L}\psi^q_R)+h.c.
\end{eqnarray}
Since in this case, the $L\leftrightarrow R$ exchange does not flip the
rotational direction, $\psi^q_L$-matter production and $\psi^q_R$-matter
production simultaneously occur, and they are exclusive to 
$\psi^q_L$-antimatter and $\psi^q_R$-antimatter production.
We thus find the net baryon number from the cosmological particle production.

On the other hand, considering the $l\leftrightarrow q$ exchange, 
$\psi^q_{L,R}$-matter production is exclusive to $\psi^l_{L,R}$-matter
production.
We thus find that the negative lepton number has to be generated
simultaneously with the positive baryon number, and vice versa.
This result is consistent with effective charge conservation.

The trick of this baryogenesis scenario is very simple in the massless limit.
If one rearranges the pairs as $\psi_1=(\psi_{1L},\psi_{1R})\equiv(\psi^q_{L},\psi^l_{R}$)
and $\psi_2=(\psi_{2L},\psi_{2R})\equiv(\psi^l_{L},\psi^q_{R}$), the Yukawa interaction can
be written as
\begin{eqnarray}
&&\left[g\phi\overline{\psi^l_R}\psi^q_L+h.c. \right]+
\left[g\phi\overline{\psi^l_L}\psi^q_R+h.c. \right]\nonumber\\
&=&
\left[g\phi\overline{\psi_{1R}}\psi_{1L}+h.c. \right]+
\left[g^*\phi^*\overline{\psi_{2R}}\psi_{2L}+h.c. \right],
\end{eqnarray}
which represents the complex Dirac mass terms for two independent Dirac
fermions $\psi_{1,2}$.
In the original $\psi^q$ and $\psi^l$, the ``exclusive pairs'' are
rearranged to avoid the cancellation of the net B and L numbers.

Remember that in the previous (Majorana) models, asymmetry vanishes for
$\boldsymbol k\rightarrow 0$.
Therefore, we are going to discuss the non-relativistic limit, where the
 original mass terms
\begin{eqnarray}
\label{eq-Diracmass-QL}
m_{q}\overline{\psi^q}\psi^q-
m_{l}\overline{\psi^l}\psi^l
\end{eqnarray}
are not negligible but $\boldsymbol k\simeq 0$ is negligible for both
fermions.
Here, the minus sign in front of $m_l$ is for later convenience.
We also assume $m_q\simeq m_l$ for simplicity.

First, consider the conventional decomposition
\begin{eqnarray}
\psi&=&\int\frac{d^3k}{(2\pi)^3}
e^{-i\boldsymbol k\cdot \boldsymbol x}\sum_s\left[
u^s_{\boldsymbol k}(t)a^s_{\boldsymbol k}
+v^s_{\boldsymbol k}(t)b^{s\dagger}_{-\boldsymbol k}\right]
\end{eqnarray}
and the single-field Dirac equation
\begin{eqnarray}
(i\slashed{\partial}-m_D)\psi&=&0.
\end{eqnarray}
Carefully following the formalism given in Ref.\cite{Peloso:2000hy}, 
one will find
\begin{eqnarray}
\dot{u}_\pm&=&ik u_{\mp}\mp i m_D u_\pm,
\end{eqnarray}
which can be written in the matrix as
\begin{eqnarray}
i\frac{d}{dt}\left(
\begin{array}{c}
u_+\\
u_-
\end{array}
\right)&=&\left(
\begin{array}{cc}
m_D& -k \\
 -k & -m_D
\end{array}
\right)
\left(
\begin{array}{c}
u_+\\
u_-
\end{array}
\right).
\end{eqnarray}

Diagonalizing this equation for constant matrix elements, 
one will find the adiabatic states (the WKB solutions) as
\begin{eqnarray}
i\frac{d}{dt}\left(
\begin{array}{c}
\tilde{u}_+\\
\tilde{u}_-
\end{array}
\right)&=&\left(
\begin{array}{cc}
\sqrt{m_D^2+k^2}& 0 \\
 0 & -\sqrt{m_D^2+k^2}
\end{array}
\right)
\left(
\begin{array}{c}
\tilde{u}_+\\
\tilde{u}_-
\end{array}
\right).\nonumber\\
\end{eqnarray}
The original $u_\pm$ and $\tilde{u}_\pm$ coincides at $k\sim 0$.

Let us introduce the interaction given in Eq.(\ref{eq-QLint})
 and take $\boldsymbol k \rightarrow 0$.
Assuming $m_q$ and $m_l$ are constant, we find 
\begin{eqnarray}
\label{eq-q+}
i\frac{d}{dt}\left(
\begin{array}{c}
u_{+}^q\\
u_{+}^l
\end{array}
\right)&=&\left(
\begin{array}{cc}
m_q & m_\Delta^* \\
m_\Delta & -m_l
\end{array}
\right)
\left(
\begin{array}{c}
u_{+}^q\\
u_{+}^l
\end{array}
\right).
\end{eqnarray}
and
\begin{eqnarray}
i\frac{d}{dt}\left(
\begin{array}{c}
u_{-}^q\\
u_{-}^l
\end{array}
\right)&=&\left(
\begin{array}{cc}
-m_q & -m_\Delta^* \\
-m_\Delta & +m_l
\end{array}
\right)
\left(
\begin{array}{c}
u_{-}^q\\
u_{-}^l
\end{array}
\right).
\end{eqnarray}
At this moment, the cancellation could not be clear from the above equations.

Decoupling the equations, we find
\begin{eqnarray}
\ddot{u}_{+}^q+\left[-\frac{\dot{m}_\Delta^*}{m_\Delta^*}+i(m_q-m_l)\right]\dot{u}_{+}^q&&\nonumber\\
+\left[|m_\Delta|^2+m_q m_l-i\frac{\dot{m}_\Delta^*}{m_\Delta^*} m_q\right]u_{+}^q&=&0
\end{eqnarray}
Redefining the interaction as $m_\Delta=m_0 e^{i\theta(t)}$, we have
\begin{eqnarray}
\frac{\dot{m}_\Delta^*}{m_\Delta^*}&=&-i\dot{\theta}.
\end{eqnarray}
Also defining $\delta M\equiv m_q-m_l\ge0$, the equation becomes
\begin{eqnarray}
&&\ddot{u}_{+}^q+i\left[\dot{\theta}+\delta M\right]\dot{u}_{1+}\nonumber\\
&&+\left[|m_\Delta|^2+ m_q m_l-\dot{\theta} m_q\right]u_{+}^q=0
\end{eqnarray}
The standard form of the EWKB will be obtained using
\begin{eqnarray}
\label{eq-st-ql}
\hat{u}_{+}^q&\equiv&\exp
 \left(\frac{i}{2}\int^t\left[\dot{\theta}+\delta M\right]dt\right)u_{+}^q.
\end{eqnarray}
This also removes the time-dependence of the off-diagonal elements of
the Landau-Zener formalism.
Finally, we have
\begin{eqnarray}
&&\ddot{\hat{u}}_{+}^q+\left[
-\frac{i}{2}\ddot{\theta}+\frac{3(\dot{\theta}+\delta M)^2}{4}\right.\nonumber\\
&&\left.+m_0^2+m_l
m_q-\dot{\theta} m_q
\right]u_{+}^q\nonumber\\
&=&0.
\end{eqnarray}
It seems difficult to understand the situation from the decoupled
equations.
Therefore, we are going back to the original matrix form.

If Eq.(\ref{eq-st-ql}) has been applied to Eq.(\ref{eq-q+}), one will find
\begin{eqnarray}
i\frac{d}{dt}\left(
\begin{array}{c}
\hat{u}_{+}^q\\
\hat{u}_{+}^l
\end{array}
\right)&=&\left(
\begin{array}{cc}
\frac{M_{1/2} -\dot{\theta}}{2} & m_0 \\
m_0 & -\frac{M_{1/2}-\dot{\theta}}{2}
\end{array}
\right)
\left(
\begin{array}{c}
\hat{u}_{+}^q\\
\hat{u}_{+}^l
\end{array}
\right),
\end{eqnarray}
and
\begin{eqnarray}
i\frac{d}{dt}\left(
\begin{array}{c}
\hat{u}_{-}^q\\
\hat{u}_{-}^l
\end{array}
\right)&=&\left(
\begin{array}{cc}
-\frac{M_{1/2} +\dot{\theta}}{2} & -m_0 \\
-m_0 & \frac{M_{1/2}+\dot{\theta}}{2}
\end{array}
\right)
\left(
\begin{array}{c}
\hat{u}_{-}^q\\
\hat{u}_{-}^l
\end{array}
\right),
\end{eqnarray}
where $M_{1/2}\equiv (m_q+m_l)/2$.
This form is very familiar in this paper.
From the above equations, we found that the situation is quite similar
to the single Dirac fermion, and the matter-antimatter
asymmetry is indeed canceled (i.e, net baryon and lepton numbers 
vanish) in the non-relativistic limit.

For completeness, let us consider a generalization of the scenario.
Since the left-handed and the right-handed fermions are independent
particles, one can also consider the Yukawa interactions written as
\begin{eqnarray}
\left[m_{\Delta ij} \overline{\psi_{Lj}}\psi_{Ri}
 +h.c.\right]+\left[m_{\Delta ji} \overline{\psi_{Li}}\psi_{Rj}
 +h.c.\right],
\end{eqnarray}
where $m_{\Delta ij}\ne m_{\Delta ji}$ is possible.

\section{Conclusions and discussions}

As we have seen explicitly, the origin of the
asymmetry in the particle production from the rotational motion is
essentially the exclusive particle production.
This property has been found by using the Landau-Zener model and 
the EWKB Stokes lines for the decoupled equations.
We also pointed out that the perturbative expansion may sometimes
change the Stokes lines.
To avoid such discrepancy, one has to draw the Stokes lines of the
original theory first and consider perturbation at the points where the
Stokes lines cross the real-time axis.
For the Dirac fermion, because of the left and right-handed components,
 the total asymmetry cancels for the single-field model.
For the multiple Dirac fermions, we have seen that cancellation can be
avoided in the relativistic limit.
In all cases, the asymmetry vanishes for $k\rightarrow 0$.

Perturbative expansion is sometimes used before non-perturbative
calculation.
Our concern was that such simplification may change the
structure of the Stokes lines of the original theory and the theory
after perturbative expansion cannot reproduce the physics of the
original theory.
For physics, the two theories could not always be needed to be identical,
 but the difference has to be under control.
To avoid the problem, one has to draw first the Stokes line of the
original theory to find a separable structure of the Stokes lines,
where local expansion is applicable.
Indeed, the idea described in Zener's original paper\cite{Zener:1932ws}
follows this regulation.
This strategy is crucial for a damped oscillation,
as is shown in Fig.\ref{fig_exactpm}. 

We started with the simplest example (i.e, simple
rotation $\Delta(t)=Ae^{i\omega_0 t}$),
and carefully examined the essential mechanism of the asymmetry
production for typical situations, comparing the structure of the Stokes
lines and the original equations.

\section{Acknowledgment}
The authors would like to thank Nobuhiro Maekawa for collaboration in
the very early stages of this work.
SE  was supported by the Sun Yat-sen University Science Foundation.

\appendix
\section{A short introduction to the EWKB and the Borel resummation}
\label{app-srevEWKB}
As we have mentioned in the introduction of this paper and in 
our previous paper\cite{Enomoto:2020xlf}, there has been many papers in
which the Stokes phenomena is used to study particle production.
However, we feel that the crucial difference between the conventional
WKB approximation and the exact WKB is not well understood widely.

First, the conventional WKB approximation generically gives a divergent
power series, while the EWKB considers the Borel resummation to control
the WKB expansion.

Second, the following misinterpretation could arise.
Since the above situation is drastically improved by taking the Borel
resummation, one might naively think that the conventional WKB
is certificated by the EWKB, and there is no crucial
diference at the end.
This speculation is sometimes very dangerous.

The discrepancy will be obvious when $Q(x)$ of the ``Schr\"odinger
equation'' 
\begin{eqnarray}
\left[-\frac{d^2}{dx^2}+\eta^2 Q(x)
\right]\psi(x,\eta)&=&0, \label{eq:Schrodinger_eta}
\end{eqnarray}
is replaced by $Q(x,\eta)$.
In this case, the discussion of the stokes phenomena becomes quite vague
(or at most very complicated) for the conventional WKB\cite{Berry:1972na}. 

Now remember the model discussed in Sec.\ref{subsec-lin}.
Although both our numerical calculation and the perturbative
calculation of Ref.\cite{Dolgov:1996qq} are giving 
the correct answer, it would be difficult for the conventional WKB to explain
why the asymmetry of the Stokes lines disappears for finite $\hbar$. 
Using the EWKB, one can easily understand that the Stokes lines have
to be calculated for $S_{-1}$, where $S_{-1}$ is the coefficient
of $\eta$ of the WKB expansion.
For the conventional WKB, the asymmetry of the Stokes lines
seems to disappear only in the limit $\hbar\rightarrow 0$, and 
something other than the conventional WKB is required to explain the situation.

In the followings, we show explicitly what determines the Stokes phenomena in the
EWKB and why the Borel resummation is crucial for the calculation.

\subsection{What is the Borel resummation?}
Let us solve the very simple Ordinary Differential Equation(ODE)
\begin{eqnarray}
z^2 \frac{du}{dz}&=&-u+z, \,\,\, u(0)=1
\end{eqnarray}
using expansion into power series
\begin{eqnarray}
u(z)&=&\sum_{n=0}^{\infty}u_n z^n.
\end{eqnarray}
One can find the solution
\begin{eqnarray}
u(z)&=&\sum_{n=0}^\infty (-1)^n n! z^n+1,
\end{eqnarray}
but this is a divergent power series because of the factor $n!$.
To remedy the situation, we note that the Laplace transformation 
of a function $f$
\begin{eqnarray}
(Lf)(t)&=&\int^\infty_0 e^{-t \tau} f(\tau) d\tau
\end{eqnarray}
gives (after partial integration)
\begin{eqnarray}
L\left[\frac{df}{d\tau}\right]&=&t(Lf)-f(0),
\end{eqnarray}
which can be applied to $f=\tau^n/n!$ to give
\begin{eqnarray}
L\left[\frac{\tau^n}{n!}\right]&=& t^{-(n+1)}.
\end{eqnarray}
Since the factor $1/n!$ arises in this formula, it would be natural to
expect that after using the inverse Laplace transformation, the factor
$n!$ of the divergent power series can be removed.
This speculation is correct.

Replacing $z$ by $t\equiv 1/z$, and after executing the inverse Laplace
transformation, one will find 
\begin{eqnarray}
(L^{-1} u)(\tau)&=& \sum_{n=0}^\infty (-1)^n
 n!\left(\frac{\tau^n}{n!}\right)\nonumber\\
&=&\frac{1}{1+\tau}.
\end{eqnarray}
After the Laplace transformation, the original power series can be
written by the integral
\begin{eqnarray}
u_B(z)&=&\int^\infty_0e^{-\tau/z}\left(\frac{1}{1+\tau}\right)d\tau.
\end{eqnarray}
This is the basic strategy of the Borel resummation.
After defining $s\equiv \tau/z$, $U(z)$ is given by
a simpler form
\begin{eqnarray}
u_B(z)&=&\int^\infty_0e^{-s}\left(\frac{1}{s+z^{-1}}\right)ds.
\end{eqnarray}
If the pole at $s=z^{-1}$ moves around the origin, it
crosses the integration contour and picks up a residue to give an
additional contribution $2\pi i e^{1/z}$ (the Stokes phenomena).

Application of this simple idea to the WKB expansion is not so much
simple but the calculation is straight.
Let us see what happens in the WKB expansion and how one can understand
the origin of the Stokes phenomena in terms of the Borel resummation.

\subsection{The Borel resummation for the WKB expansion}
Our starting point is the
``Schr\"odinger equation'' in quantum mechanics given by 
\begin{eqnarray}
\left[-\frac{d^2}{dx^2}+\eta^2 Q(x,\eta)
\right]\psi(x,\eta)&=&0,
\end{eqnarray}
where 
\begin{eqnarray}
Q(x,\eta)&=&V-E
\end{eqnarray}
for the potential $V$ and the energy $E$.
Just for simplicity, we assume that $Q$ is given by
\begin{equation}
 Q(x,\eta)=\sum_{k=0}^\infty\eta^{-2k}Q_{2k}(x), \label{eq:Q_eta_series}
\end{equation}
which simplifies the calculation as is shown in Eq.(\ref{eq:WKB_sol_odd}) later.

Assuming that the solution $\psi$ is written by
$\psi(x,\eta)=e^{R(x,\eta)}$, one will have 
\begin{eqnarray}
\psi&=&e^{\int^x_{x_0}S(x,\eta)dx}
\end{eqnarray}
for $S(x,\eta)\equiv \partial R/\partial x$.
We have the condition for $S$ given by
\begin{eqnarray}
-\left(S^2 +\frac{\partial S}{\partial x}\right)+\eta^2 Q&=&0.
\end{eqnarray}
Expanding $S$ as $S(x,\eta)=\sum_{n=-1}^{n=\infty}\eta^{-n} S_{n}$,
one will have
\begin{eqnarray}
S=\eta S_{-1}(x)+ S_0(x)+\eta^{-1}S_1(x)+...,
\end{eqnarray}
which gives
\begin{eqnarray}
S_{-1}^2 &=& Q_0
\end{eqnarray}
and
\begin{eqnarray}
2S_{-1}S_{2j}&=&-\left[\sum_{k\ge 0}^jS_{2k}S_{2(j-k)-1}
 +\sum_{k\ge 0}^{j-1}S_{2k+1}S_{2(j-k-1)}\right. \nonumber \\
 & & \qquad \left. + \frac{dS_{2j-1}}{dx}\right]\\
2S_{-1}S_{2j+1}&=&-\left[\sum_{k\ge 0}^jS_{2k}S_{2(j-k)}
 +\sum_{k\ge 0}^{j-1}S_{2k+1}S_{2(j-k)-1}\right. \nonumber \\
 & & \qquad \left. + \frac{dS_{2j}}{dx}-Q_{2(j+1)}\right]
\end{eqnarray}
for $j\geq 0$.  Note that the sign dependence of $S_{-1}=\pm\sqrt{Q_0}$ reflects to the odd terms as
$\pm S_{2j+1}$ but not to the even terms.
Using this relation, one will have a simpler form
\begin{eqnarray}
\psi&=&\frac{1}{\sqrt{S_{odd}}}e^{\int^x_{x_0}S_{odd}dx}\\
&&S_{odd}\equiv\sum_{j\ge 0}\eta^{1-2j}S_{2j-1}.
\end{eqnarray}
Depending on the sign of the first $S_{-1}$, there are
two solutions $\psi_\pm$, which are given by
\begin{eqnarray}
\psi_{\pm}&=&\frac{1}{\sqrt{S_{odd}}}\exp\left(\pm \int^x_{x_0}S_{odd} dx\right). \label{eq:WKB_sol_odd}
\end{eqnarray}
The above WKB expansion is usually divergent.

\begin{figure}[t]
\centering
\includegraphics[width=1.0\columnwidth]{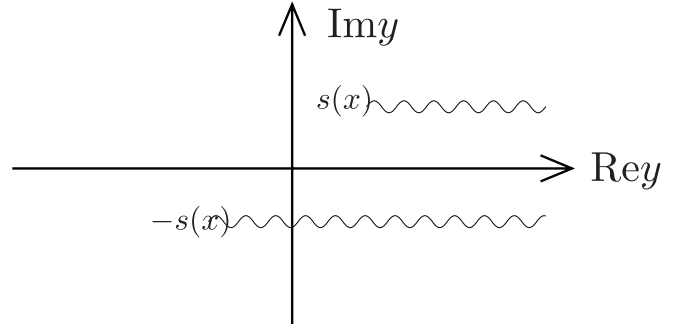}
 \caption{The wavy lines are the paths of integration in the Borel
 resummation.
Since the paths overlap when $\mathrm{Im}[s]=0$, the Stokes phenomena
 appears if one crosses the Stokes lines defined by $\mathrm{Im}[s]=0$.}
\label{fig_borelintegral}
\end{figure}

Let us consider the Borel resummation for the WKB expansion.
Since $x$ appears in the coefficients of the $\eta$ expansion, we
consider the Borel resummation for $\eta$, not for $x$. 
Note that this point is different from the simplest example discussed in the
previous section.
We thus have
\begin{eqnarray}
\psi_\pm &\rightarrow&\Psi_\pm\equiv\int^\infty_{\mp s(x)}e^{-y\eta}\psi_\pm^B(x,y)dy,\\
&&s(x)\equiv \int^x_{x_0}S_{-1}(x)dx,
\end{eqnarray}
where the $y$-integral is parallel to the real axis.
What is important here for the Stokes phenomena is that the starting point of
the integration is naturally moved from the origin to $s(x)$ due to
$S_{-1}\eta$ and $y\eta$ in the exponent.
Now the starting points of the integration paths depend
explicitly on the original coordinate $x$.
See Fig.(\ref{fig_borelintegral}) for more information about the contour
and the Stokes phenomena.

Therefore, one can easily expect that if one moves in the complex $x$
space, the integration contour may cause the Stokes phenomena similar to
the one discussed in the previous section.

See also the simplest example in Sec.\ref{subsec-lin}, which clearly 
explains why EWKB is useful.

Locally, the Stokes phenomenon around the turning points ($Q=0$) can
 also be explained by using the Airy function ($Q(x)=x$).
However, to generalize the discussion, the technique of
the Borel resummation seems to be inevitable.

\section{The Exact WKB for cosmological particle production}
\label{app-reviewEWKB}
In this appendix, we show the calculation of cosmological particle
production in the light of the Exact WKB.
We show explicitly the Stokes lines for bosonic and fermionic preheating
and examine the adequacy of the conventional linear approximation at the
enhanced symmetry point (ESP).
After the linear approximation, which is local, the connection formula
at that point is identical to the scattering problem by an inverted
quadratic potential (for bosonic preheating) or the Landau-Zener
transition (for fermionic preheating),  
whose Stokes lines are given by the Merged pair of simple Turning Points(MTP).
If the imaginary part of $Q_0$ remains, the MTP is unmerged. 

The Stokes lines in this appendix are drawn by using
Mathematica\cite{Mathematica-stokes}.

\subsection{Bosonic preheating}
We first review the standard calculation based on the special
function and then explain the EWKB calculation.

\subsubsection{Standard calculation : 
Exact local solution after linear
   approximation, $m(t)\rightarrow m(t_i) + \dot{m}(t_i)(t-t_i)$}
Just after inflation, the motion of the inflaton field is usually a damped
oscillation, whose particle production cannot be solved exactly.
However, at least near the center of the oscillation, or very close to
the ESP where particle production is likely to take place, the linear
approximation can be made with respect to $t$.
We are going to review the local solution obtained from the linear
   approximation at the ESP.

One can rewrite the inflaton motion as 
$\phi(t)=\phi(t_i)+\dot{\phi}(t_i)(t-t_i)$,
where $t_i$ is the $i$-th minimum of $m^2(t)$.
Typically, the mass of a scalar field (e.g, $\chi$) is 
given by 
\begin{eqnarray}
m^2_\chi(t)&=&m_0^2 +g^2_2 \phi(t)^2,
\end{eqnarray}
where $\phi(t)$ is the oscillating inflaton field.
If we consider the Lagrangian
\begin{eqnarray}
{\cal L}_\chi&=&\frac{1}{2}\partial_\mu \chi\partial^\mu \chi
 -\frac{1}{2}m_0^2\chi^2 -\frac{1}{2}g_2^2 \phi(t)^2\chi^2,
\end{eqnarray}
the equation of motion is 
\begin{eqnarray}
\frac{d^2 \chi}{dt^2}+\left[k^2+m^2_\chi(t)\right]\chi=0.
\end{eqnarray}
If one replaces $\phi(t)$ with $\phi(t)\simeq vt$, the equation is
equivalent to the Schr\"odinger equation with the
``inverted quadratic potential''.
Then the connection formula is obtained by solving the
quantum scattering problem.
The ``potential'' is given by
\begin{eqnarray}
V(t)&=&-\left(g_2^2v^2\right)t^2,
\end{eqnarray}
where the corresponding ``energy'' is 
\begin{eqnarray}
E&=&k^2+m_0^2.
\end{eqnarray}
Since $E>V$ is always true in this case, there is no classical turning
point of the scattering problem.
Here we have neglected the expansion of the Universe, which can be
introduced by making redefinitions of the parameters.

Typically, the WKB expansion is used to find
\begin{eqnarray}
\label{eq-WKB}
\chi_k(t)&=& \frac{\alpha_k(t)}{\sqrt{2\omega_k}}e^{-i \int^t \omega dt}
+\frac{\beta_k(t)}{\sqrt{2\omega}}e^{+i \int^t \omega dt},
\end{eqnarray}
where
\begin{eqnarray}
\omega_k(t)&\equiv&k^2+m_\chi^2(t).
\end{eqnarray}
We take $\alpha_k=1, \beta_k=0$ for the initial vacuum state.
The distribution of the particle in the final state is 
\begin{eqnarray}
n_\chi(k)=|\beta_k|^2,
\end{eqnarray}
which can be found by solving the scattering problem of the
corresponding Schr\"odinger equation.
For the above model (i.e, scattering by the inverted quadratic
potential), the following Weber equation
\begin{eqnarray}
y''(z)+\left(\nu+\frac{1}{2}-\frac{1}{4}z^2\right)y(z)=0
\end{eqnarray}
has the solution $D_\nu(z), D_{-\nu-1}(iz)$.
More specifically, one can define
\begin{eqnarray}
z&\equiv& ie^{i\pi/4}\sqrt{2g_2v}t
\end{eqnarray}
in the original field equation to find 
\begin{eqnarray}
\frac{d^2 \chi}{dz^2}+\left[\nu+\frac{1}{2} -\frac{1}{4}z^2\right]\chi=0.
\end{eqnarray}
Here we defined
\begin{eqnarray}
\nu=\frac{k^2+m_0^2}{2g_2v}i-\frac{1}{2},
\end{eqnarray}
and for later use we define 
\begin{eqnarray}
\kappa&\equiv& \frac{k^2+m_0^2}{2g_2v}
\end{eqnarray}
and 
\begin{eqnarray}
\nu=i\kappa -\frac{1}{2}.
\end{eqnarray}
Here, $\kappa$ is later used to estimate the particle production.
Comparing the asymptotic forms of the Weber function, one will
find\cite{Enomoto:2020xlf}
\begin{eqnarray}
D_\nu(z)&\simeq&
e^{-i\frac{g_2v}{2}t^2}e^{(i\kappa-\frac{1}{2})\left(\log(\sqrt{2g_2 v}
|t|)-i\frac{\pi}{4}\right)},\nonumber\\
D_{-\nu-1}(iz)&\simeq& e^{+i\frac{g_2v}{2}t^2}e^{(-i\kappa-\frac{1}{2})\left(\log(\sqrt{2g_2 v}
						 |t|)+i\frac{\pi}{4}\right)}.\nonumber\\
\end{eqnarray}

In the limit $t\rightarrow -\infty$, the above solutions are
giving the $\pm$ WKB solutions of Eq.(\ref{eq-WKB}).
Therefore, one will have in the  $t\rightarrow -\infty$ limit
\begin{eqnarray}
\chi_- &\rightarrow&D_{\nu}(z)\\
\chi_+ &\rightarrow&D_{-\nu-1}(iz).
\end{eqnarray}

In the opposite $t\rightarrow +\infty$ limit, we have
\begin{eqnarray}
D_\nu(z)&\simeq& e^{-i\frac{g_2v}{2}t^2}e^{(i\kappa+\frac{1}{2})
\left(\log(\sqrt{2g_2 v}t)+i\frac{3\pi}{4}\right)}\nonumber\\
&&+i\frac{\sqrt{2\pi}}{\Gamma(-\nu)}e^{i\frac{g_2v}{2}t^2}
e^{-\kappa \pi}e^{(-i\kappa-\frac{1}{2})\left(\log(\sqrt{2g_2 v}t)
+i\frac{3\pi}{4}\right)},\nonumber\\
\end{eqnarray}
which shows that in the $t=+\infty$ limit the asymptotic
form of the exact solution $D_\nu(z)$ is the mixture of the
$\pm$ WKB solutions.
Thus we find the exact connection formula using the special function.

In this case, the connection formula gives the Bogoliubov
transformation of the WKB solutions.\footnote{One has to use
\begin{eqnarray}
\Gamma(z)\Gamma(1-z)&=&\frac{\pi}{\sin \pi z}\\
\Gamma(\bar{z})&=&\overline{\Gamma(z)}\\
1+\nu&=&1+\left(i\kappa-\frac{1}{2}\right)=-\overline{\nu}
\end{eqnarray}
for the calculation of $\Gamma(-\nu)=\Gamma(-i\kappa+\frac{1}{2})$.}
Finally, we obtain the connection matrix
\begin{eqnarray}
\label{eq-MTP-connect}
&&\left(
\begin{array}{c}
\alpha_k^{+\infty}\nonumber\\
\beta_k^{+\infty}
\end{array}
\right)\\
&=&
\left(
\begin{array}{cc}
\sqrt{1+e^{-2\pi \kappa}}e^{i\theta_1} & ie^{-\pi\kappa+i\theta_2}\\
-ie^{-\pi\kappa-i\theta_2} &\sqrt{1+e^{-2\pi \kappa}}e^{-i\theta_1} 
\end{array}
\right)
\left(
\begin{array}{c}
\alpha_k^{-\infty}\\
\beta_k^{-\infty}
\end{array}
\right).\nonumber\\
\end{eqnarray}
Here, all the phase parameters are put into $\theta_{1,2}(k)$.

\subsubsection{The Exact WKB for bosonic preheating}
First, we use the EWKB to find the connection formula of the local
solution, which is the exact local solution of the theory after linear expansion.
Then we show the global structure of the original Stokes lines and show
why such a local solution is justified.

\begin{figure}[ht]
\centering
\includegraphics[width=0.9\columnwidth]{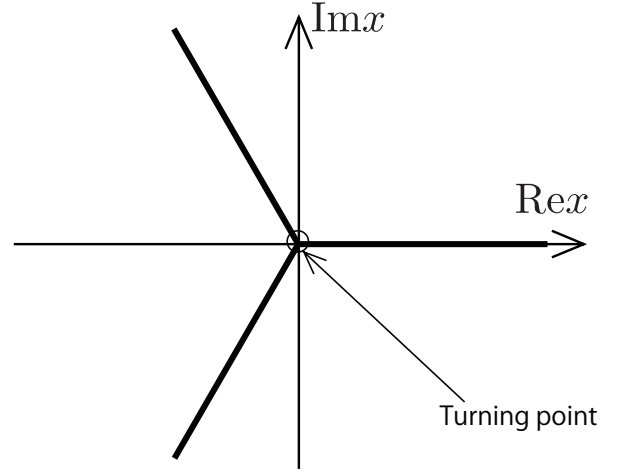}
 \caption{The Stokes lines for $Q=x$.}
\label{fig_airy}
\end{figure}

The Stokes phenomenon around the turning points can be easily understood
using the Airy function ($Q(x)=x$).
In this case, the Stokes lines are defined by
\begin{eqnarray}
\mathrm{Im} [s(x)]&=& \mathrm{Im} \left[\int^x_0 x^{1/2}dx \right]\nonumber\\
&=&\mathrm{Im} \left[\frac{2}{3}x^{3/2}\right]=0,
\end{eqnarray}
which is shown in Fig.\ref{fig_airy}.
If $\mathrm{Re}[s(x)]>0$, $\psi_+$ is dominant on the Stokes line, while
if $\mathrm{Re}[s(x)]<0$, $\psi_-$ is dominant.

When the integration paths of the Borel resummation overlap on the Stokes line, it may develop
additional contributions, which is called the Stokes phenomenon.
We have the following connection formulae:
\begin{itemize}
\item Crossing the $\psi_+$-Dominant Stokes line with anticlockwise rotation (seen from the turning
      point)
\begin{eqnarray}
\Psi_+&\rightarrow& \Psi_+ +i\Psi_-\\
\Psi_-&\rightarrow& \Psi_-
\end{eqnarray}
\item Crossing the $\psi_-$-Dominant Stokes line with anticlockwise rotation (seen from the turning
      point)
\begin{eqnarray}
\Psi_-&\rightarrow& \Psi_- +i\Psi_+\\
\Psi_+&\rightarrow& \Psi_+
\end{eqnarray}
\item Inverse rotation gives a minus sign in front of $i$.
\end{itemize}
Here, we are confined to the Airy function, but generalization of this
idea requires more calculation\cite{Virtual:2015HKT}. 

\begin{figure}[t]
\centering
\includegraphics[width=0.6\columnwidth]{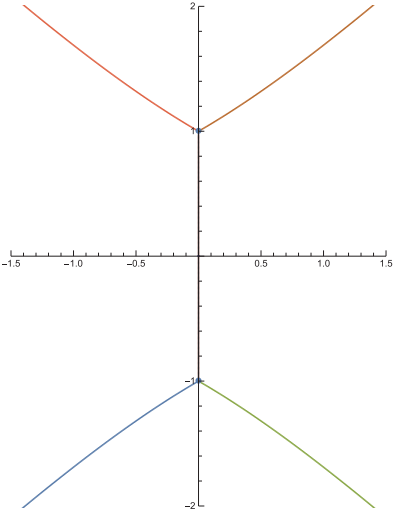}
 \caption{The Stokes lines for the inverted quadratic potential
 ($Q=-1-t^2$). }
\label{fig_stokes_invquad}
\end{figure}

Let us use these simple formulae to solve the scattering problem 
by the inverted quadratic potential ($E>0$).
We show the Stokes lines in Fig.\ref{fig_stokes_invquad}, which has
a typical MTP structure.

\begin{figure}[t]
\centering
\includegraphics[width=0.6\columnwidth]{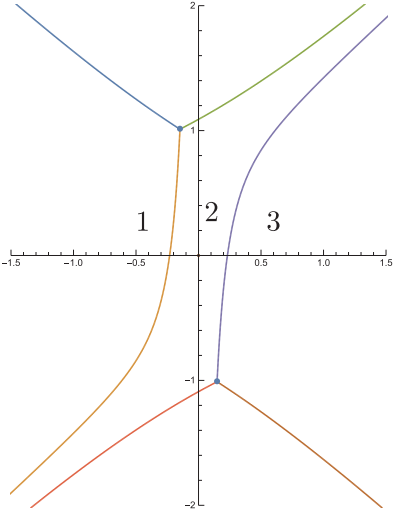}
 \caption{The Stokes lines of the MTP are unmerged by $[Q=-1-t^2 ]\rightarrow [Q=-(1 +0.3 i)-t^2] $.}
\label{fig_stokes_invquad-deltae}
\end{figure}

The degeneracy of the Stokes line can be solved (the MTP can be
unmerged) by introducing imaginary
parameters, as is explicitly shown in Fig.\ref{fig_stokes_invquad-deltae}.
One can calculate the connection formula
$1\rightarrow2\rightarrow3$, as is shown in Ref.\cite{Enomoto:2020xlf}.
One thing that is not trivial is the relation between the normalization
factor and the gap, which appears when the sign of the imaginary part is
changed (e.g, $E+i \epsilon\leftrightarrow E-i\epsilon$)
For the normalization factor, we have 
\begin{eqnarray}
\psi_{\pm}&=&\frac{1}{\sqrt{S_{odd}}}
\exp\left(\pm \int^x_{x_0}S_{odd} dx\right)\nonumber\\
&=&\psi^{(\infty)}_\pm 
\exp\left(\pm \int^x_{x_0}\left(S_{odd}-\eta S_{-1}\right) dx\right),\\
\psi^{(\infty)}_\pm&=&\frac{1}{\sqrt{S_{odd}}}
\exp\left(\pm \int^x_{x_0}S_{-1} dx\right) \nonumber\\
&&\times
\exp\left(\pm \int^x_{\infty}\left(S_{odd}-\eta S_{-1} \right)dx\right).
\end{eqnarray}
Among them, what is not trivial is 
\begin{eqnarray}
\exp\left(\pm \int^x_{x_0}\left(S_{odd}-\eta S_{-1}\right) dx\right).
\end{eqnarray}
This gives for $Q(x)=\lambda-\frac{x^2}{4}$,
\begin{eqnarray}
&&2\int^x_{2\sqrt{\lambda}}\left(S_{odd}-\eta S_{-1}\right) dx\nonumber\\
&=&\sum^\infty_{n=1} \frac{2^{1-2n}-1}{2n(2n-1)}B_{2n}(i\eta\lambda)^{1-2n},
\end{eqnarray}
where $B_{2n}$ is the Bernoulli number.
Because of this, one has the gap given by
\begin{eqnarray}
\Psi_+^{(\mathrm{Im}\lambda<0)}&=&\sqrt{1+e^{-2\pi\lambda \eta}}
\Psi_+^{(\mathrm{Im}\lambda>0)}.
\end{eqnarray}
The calculation can be generalized to give 
the factor appearing on both sides of more generic MTP\cite{Silverstone:2008,
Aoki:2009}.

\begin{figure}[t]
\centering
\includegraphics[width=0.8\columnwidth]{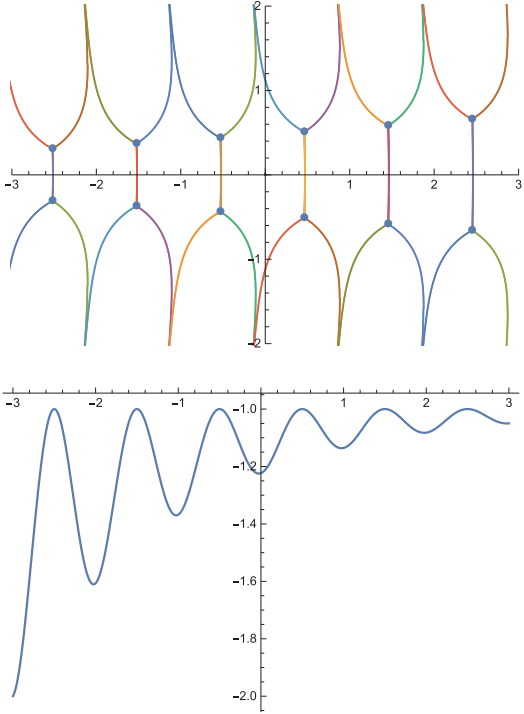}
 \caption{The Stokes lines for $Q(t)=-1-e^{-0.3(t+3)}\cos^2(2t)$ and a plot of
 $Q(t)$. Separable MTP structures are appearing at the minima of $m^2_\chi$}
\label{fig_stokes_scalardampedosc}
\end{figure}

From the above results, one can understand that each separable MTP structure
will have the local connection matrix given by Eq.(\ref{eq-MTP-connect}).
This idea is very useful for calculating particle production by a
damped oscillation.
To understand more about the idea, we need to draw the global structure
of the Stokes lines.
We assume that the mass is given by $m_\chi(t)^2=g^2 \phi(t)^2 = g^2
(e^{-\Gamma t}\cos(a t))^2$.
From Fig.\ref{fig_stokes_scalardampedosc}, one can easily understand
that the linear expansion at the local minimum of $m_\chi^2$ (i.e, the
local maximum of the ``potential'') is giving a good
approximation.

As we have stated above, the MTP structure is obvious in bosonic
preheating.
Because of the separable MTP structure, linear approximation at the ESP
is easily justified and the calculation becomes local.
The situation is slightly complicated for fermionic preheating.

\subsection{Fermionic preheating}
As we have described for bosonic preheating, we start with the local
solution after linear approximation, which is nothing but the
Landau-Zener model.
Since the fermionic preheating with a linear $\phi(t)$ is an old idea,
we carefully follow Ref.\cite{Peloso:2000hy} to avoid confusion.

We consider the Dirac fermion whose mass is 
\begin{eqnarray}
m_D(t)&=& m_0 + g\phi(t),
\end{eqnarray}
where $m_D$ is real.
The Dirac equation is 
\begin{eqnarray}
(i\slashed{\partial}-m_D)\psi&=&0,
\end{eqnarray}
whose solution can be decomposed as
\begin{eqnarray}
\psi&=&\int\frac{d^3k}{(2\pi)^3}
e^{-i\boldsymbol k\cdot \boldsymbol x}\sum_s\left[
u^s_{\boldsymbol k}(t)a^s_{\boldsymbol k}
+v^s_{\boldsymbol k}(t)b^{s\dagger}_{-\boldsymbol k}\right].
\end{eqnarray}
Following Ref.\cite{Peloso:2000hy}, we choose the momentum along the
third direction $k=k_z$ and introduce $u_\pm$.
Then one obtains a two-component differential
equation\footnote{According to \cite{Peloso:2000hy},
the representation of the gamma matrices is chosen as
\begin{equation}
 \gamma^0=\left(\begin{array}{cc}\mathbf{1} & \\ & -\mathbf{1} \end{array} \right), \quad
 \gamma^1=\left(\begin{array}{cc} & -i\sigma^2 \\ -i\sigma^2 & \end{array} \right),
\end{equation}
\begin{equation}
 \gamma^2=\left(\begin{array}{cc} & i\sigma^1 \\ i\sigma^1 & \end{array} \right), \quad
 \gamma^3=\left(\begin{array}{cc} & \mathbf{1} \\ -\mathbf{1} & \end{array} \right).
\end{equation}}
\begin{eqnarray}
\label{eq-F-EOM}
\dot{u}_\pm&=&ik u_{\mp}\mp i m_D u_\pm,
\end{eqnarray}
which can be written as
\begin{eqnarray}
i\frac{d}{dt}\left(
\begin{array}{c}
u_+\\
u_-
\end{array}
\right)&=&\left(
\begin{array}{cc}
m_0+g\phi(t)& -k \\
 -k & -m_0-g\phi(t)
\end{array}
\right)
\left(
\begin{array}{c}
u_+\\
u_-
\end{array}
\right).\nonumber\\
\end{eqnarray}
Note that this equation is similar to the Landau-Zener model.
For $m_0+g\phi(t)= vt$, the equation of motion is nothing but the
original Landau-Zener model, which is solved in Ref.\cite{Zener:1932ws}.
Decoupling the equations, one can use the EWKB.

For the bosonic preheating scenario, we have seen that the global
structure of the original Stokes lines has separable MTP at each minimum
of $m_\chi^2$, which supports the conventional local calculation of the connection
formula\cite{Kofman:1997yn}.

\begin{figure}[t]
\centering
\includegraphics[width=1.0\columnwidth]{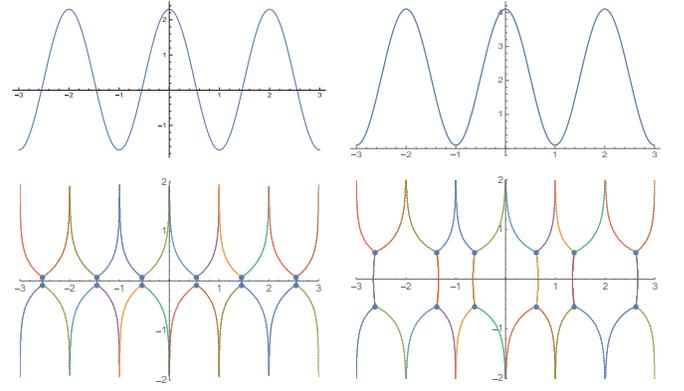}
 \caption{Left: Plot of $m_D(t)$ and the Stokes lines.
One can see that the MTP appears at each massless point (when the states cross
 in the light of the Landau-Zener model).
Right: $m_D$ cannot be massless because of large $m_0$.
Since the local linear approximation is not valid at the minimum of
 $m_D$, the corresponding scattering problem is not for the quadratic
 potential, but for the (higher) quartic potential.
One can see that the Stokes lines are giving the
 structure of the quartic scattering\cite{Enomoto:2020xlf} at each minimum.}
\label{fig_FPR-stokes1}
\end{figure}

Let us see if such structure appears in fermionic preheating.
Decoupling the equations, we find for $u_+$;
\begin{eqnarray}
u_+&=& -\left(\frac{k^2+m_D^2}{\hbar^2}+i\frac{\dot{m}_D}{\hbar}\right)u_+.
\end{eqnarray}
For $m_D(t)=m_0 + A \cos\omega t$, we find
\begin{eqnarray}
Q_0&=&-k^2-(m_0 + A \cos\omega t)^2.
\end{eqnarray}
We show the ``potential'' $V(t)=-k^2-(m_0 + A \cos\omega t)^2$
and the Stokes lines in Fig.\ref{fig_FPR-stokes1}, which shows the
expected MTP structure at each massless point, where 
``the states cross'' in terms of the Landau-Zener model. 
We thus confirmed that the linear expansion at the crossing point, which is used in
the Landau-Zener model, is justified because of the separable structure of
the Stokes lines.

\begin{figure}[t]
\centering
\includegraphics[width=1.0\columnwidth]{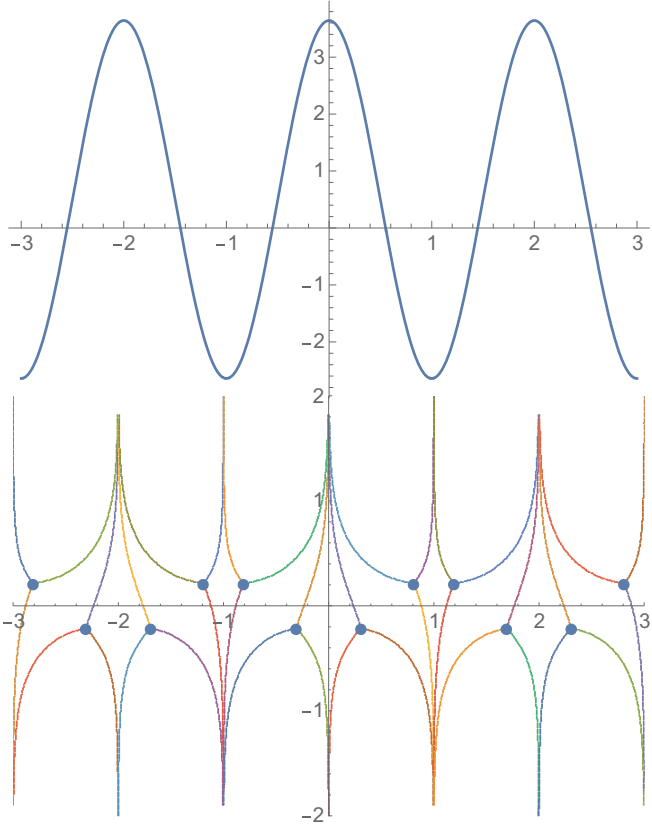}
 \caption{A plot of $m(t)$ and the Stokes lines when $\hbar^{-1}$ is
 explicit in the phase of the periodic function. One can see that the
 MTP is unmerged by the imaginary part.} 
\label{fig_FPR-stokes2}
\end{figure}

On the other hand, if one chooses $m(t)=m_0 + A \cos(a t/\hbar)$, in
which $\hbar$ is explicit, one will find an additional
imaginary part in
\begin{eqnarray}
Q_0&=&-k^2-\left(m_0 + A \cos\frac{a
				    t}{\hbar}\right)^2
-i aA \sin\frac{a t}{\hbar},\nonumber\\
\end{eqnarray}
which unmerges the MTP.
The Stokes lines are shown in Fig.\ref{fig_FPR-stokes2}, which 
have the expected structure.

On the other hand, if one expands this (latter) $m_D$ at the massless
point ($t=t_i$), one will have 
\begin{eqnarray}
m_D&\simeq&m_0+ \frac{1}{\hbar}\left[A a\sin\frac{a t_i}{\hbar}\right](t-t_i),
\end{eqnarray}
which formally ruins the WKB expansion since it contains an extra
 $\hbar{^{-1}}$ in the ``small'' part.
In this sense, it would be better to define $\omega\equiv a/\hbar$ to
control the $\hbar$ expansion before discussing the local (linear) expansion.

\begin{figure}[t]
\centering
\includegraphics[width=1.0\columnwidth]{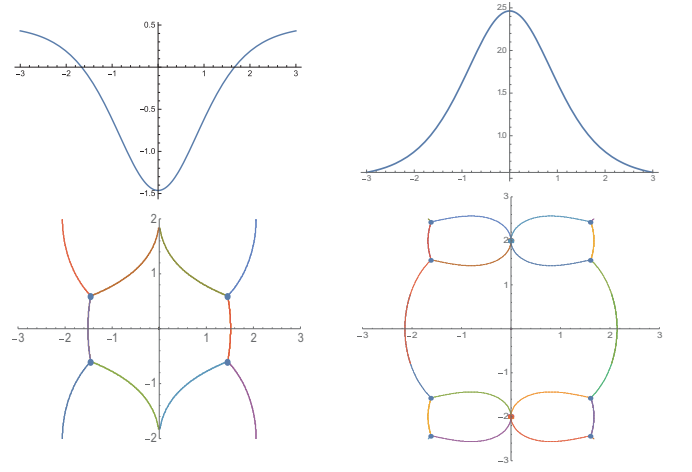}
 \caption{Left: Plot of $m_D(t)$ and the Stokes lines when $m_D$ has
 $m_D=0$ for real $t$.
One can see that the MTP-like structure appears at each massless point.
(The Stokes lines are infinitely degenerated in this case.)
Right: Although $m_D$ cannot have $m_D=0$ for real $t$, the Stokes lines are
 crossing the real $t$ axis.
This suggests that particle production may take place when the velocity
 increases.} 
\label{fig_FPR-stokes3}
\end{figure}

Finally, let us see what happens if $m_D(t)$ has a bump;
\begin{eqnarray}
m_D(t)&=& m_0  \pm\frac{A e^{\gamma t}}{\left(1+e^{\gamma t}\right)^2}.
\end{eqnarray}
In this case, the massless points can appear only for the minus sign.
We show the Stokes lines in Fig.\ref{fig_FPR-stokes3}, which shows the
expected MTP-like structure on the left.
On the right, however, the Stokes phenomena seem to be possible although
state crossing (in the light of the Landau-Zener model) does not appear.
The particle production without state crossing can be verified by
numerical calculation, although the magnitude is suppressed.

\begin{figure}[t]
\centering
\includegraphics[width=1.0\columnwidth]{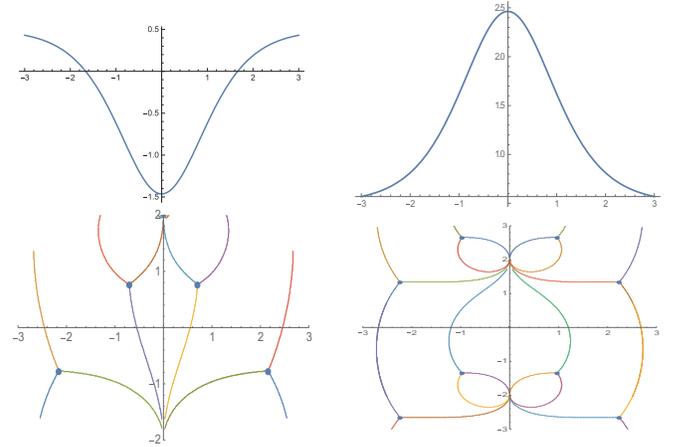}
 \caption{Left: Plot of $m_D(t)$ and the Stokes lines when $m_D$ can
 cross $m_D=0$.
Right: When $m_D$ cannot cross $m_D=0$.
The Stokes lines split due to the imaginary part.} 
\label{fig_FPR-stokes4}
\end{figure}

Also in this case, the introduction of $\hbar$ in the exponents causes a split
of the MTP-like structure.
For example, one can consider
\begin{eqnarray}
m_D(t)&=& m_0  \pm\frac{A e^{\frac{a t}{\hbar}}}{\left(1+e^{\frac{a
					      t}{\hbar}}\right)^2},
\end{eqnarray}
which introduces the additional imaginary part in $Q_0$.
The Stokes lines are shown in Fig.\ref{fig_FPR-stokes4}.

\section{Distribution of Majorana fermion} \label{sec_distribution_MF}
In this section, we show the formula of the distribution function
of the Majorana fermion described by the wave functions and its equation of evolution.

\subsection{Distribution formula} \label{sec_distribution_MF_1}

Using the mode expansion (\ref{eq_expansion_MF}), the Hamiltonian
can be represented as
\begin{eqnarray}
H &=& \int d^3x \left(-\psi_R^\dagger i\bar{\sigma}^i\partial_i\psi_R
+\frac{1}{2}m_R\psi_R^2+\frac{1}{2}m_R^*\psi_R^{\dagger 2} \right)\nonumber \\
&=& \int\frac{d^3k}{(2\pi)^3}\sum_{s=\pm}\frac{1}{2}\left[
E_k^s(a_{\boldsymbol k}^{s\dagger}a_{\boldsymbol k}^s
-a_{-\boldsymbol k}^sa_{-\boldsymbol k}^{s\dagger}) \right. \nonumber \\
& & \left. \qquad \qquad \qquad
+\left(F_k^s\cdot e^{i\theta_{\boldsymbol k}}a_{-\boldsymbol k}^sa_{\boldsymbol k}^s
+({\rm h.c.})\right) \right]\\
&=& \int\frac{d^3k}{(2\pi)^3}\sum_{s=\pm}\frac{1}{2}(
\begin{array}{cc} a_{\boldsymbol k}^{s\dagger} &
e^{i\theta_{\boldsymbol k}}a_{-\boldsymbol k}^s \end{array}) \nonumber \\
& & \qquad \qquad \quad \times
\left(\begin{array}{cc} E_k^s & F_k^{s*} \\ F_k^s & -E_k^s \end{array}\right)
\left(\begin{array}{c}a_{\boldsymbol k}^s \\
e^{-i\theta_{\boldsymbol k}}a_{-\boldsymbol k}^{s\dagger} \end{array}\right)
\label{eq_Hamiltonian_MF}
\end{eqnarray}
where
\begin{eqnarray}
E_k^s&\equiv&-s|{\boldsymbol k}|(|u_k^s|^2-|v_k^s|^2)
+sm_Ru_k^sv_k^{s*}+sm_R^*u_k^{s*}v_k^s, \nonumber \\ \label{eq:def_E}\\
F_k^s&\equiv&-2s|{\boldsymbol k}|u_k^sv_k^s-sm_R(u_k^s)^2+sm_R^*(v_k^s)^2.\label{eq:def_F}
\end{eqnarray}
Note that $E_k^s$ and $F_k^s$ have a nontrivial relation
\begin{equation}
(E_k^s)^2+|F_k^s|^2=\omega_k^2, \qquad \omega_k\equiv\sqrt{|{\boldsymbol k}|^2+|m_R|^2}
\end{equation}
because of the normalization of the wave functions $|u_k^s|^2+|v_k^s|^2=1.$
To obtain the number operator, we diagonalize the Hamiltonian (\ref{eq_Hamiltonian_MF})
using the (time-dependent) Bogoliubov transformation
\begin{equation}
a_{\boldsymbol k}^s=\alpha_k^s(t)\bar{a}_{\boldsymbol k}^s(t)
+\beta_k^s(t)\cdot e^{-i\theta_{\boldsymbol k}}\bar{a}_{-\boldsymbol k}^{s\dagger}(t),
\label{eq_BT_MF}
\end{equation}
where $\alpha_k^s$ and $\beta_k^s$ are the Bogoliubov ``coefficients'' that satisfy
\begin{equation}
 |\alpha_k^s|^2+|\beta_k^s|^2=1.
\end{equation}
Substituting (\ref{eq_BT_MF}) into (\ref{eq_Hamiltonian_MF}), the Hamiltonian can be represented
in terms of $\bar{a}_{\boldsymbol k}^s$-basis as
\begin{eqnarray}
H &=& \int\frac{d^3k}{(2\pi)^3}\sum_{s=\pm}\frac{1}{2}(
\begin{array}{cc} \bar{a}_{\boldsymbol k}^{s\dagger} &
e^{i\theta_{\boldsymbol k}}\bar{a}_{-\boldsymbol k}^s \end{array}) \nonumber \\
& & \qquad \qquad \quad \times
\left(\begin{array}{cc} \bar{E}_k^s & \bar{F}_k^{s*} \\ \bar{F}_k^s & -\bar{E}_k^s \end{array}\right)
\left(\begin{array}{c}\bar{a}_{\boldsymbol k}^s \\
e^{-i\theta_{\boldsymbol k}}\bar{a}_{-\boldsymbol k}^{s\dagger} \end{array}\right)
\end{eqnarray}
where
\begin{eqnarray}
\bar{E}_k^s&=&(|\alpha_k^s|^2-|\beta_k^s|^2)E_k^s-\alpha_k^s\beta_k^sF_k^s-\alpha_k^{s*}\beta_k^{s*}F_k^{s*} \\
\bar{F}_k^s&=&2\alpha_k^s\beta_k^{s*}E_k^s+(\alpha_k^s)^2F_k^s-(\beta_k^{s*})^2F_k^{s*}.
\end{eqnarray}
As $\bar{F}_k^s=0$ is chosen to diagonalize Hamiltonian, one can obtain
\begin{equation}
|\beta_k^s|^2=\frac{1}{2}\left(1-\frac{E_k^s}{\omega_k}\right),
\qquad \alpha_k^s=-\frac{\omega_k+E_k^s}{F_k^s}\beta_k^{s*}.
\end{equation}
Hence the Hamiltonian can be diagonalized as
\begin{eqnarray}
H &=& \int\frac{d^3k}{(2\pi)^3}\sum_{s=\pm}\frac{1}{2}(
\begin{array}{cc} \bar{a}_{\boldsymbol k}^{s\dagger} &
e^{i\theta_{\boldsymbol k}}\bar{a}_{-\boldsymbol k}^s \end{array}) \nonumber \\
& & \qquad \qquad\times
\left(\begin{array}{cc} \omega_k & \\ & -\omega_k \end{array}\right)
\left(\begin{array}{c}\bar{a}_{\boldsymbol k}^s \\
e^{-i\theta_{\boldsymbol k}}\bar{a}_{-\boldsymbol k}^{s\dagger} \end{array}\right) \\
&=& \int\frac{d^3k}{(2\pi)^3}\sum_{s=\pm}\omega_k
\left(\bar{a}_{\boldsymbol k}^{s\dagger}\bar{a}_{\boldsymbol k}^s-\frac{1}{2}\int d^3x\right).
\end{eqnarray}
The distribution function $n_k^s$ can be obtained by the time-dependent occupation operator
and the initial vacuum state defined by $a_{\boldsymbol k}^s|0\rangle=0$ as
\begin{eqnarray}
n_k^s(t)&=& \frac{\langle0|\bar{a}_{\boldsymbol k}^{s\dagger}\bar{a}_{\boldsymbol k}^s|0\rangle}{\int d^3x} \\
&=& |\beta_k^s|^2 \: = \: \frac{1}{2}\left(1-\frac{E_k^s}{\omega_k}\right). \label{eq:distribution_formula}
\end{eqnarray}

\subsection{Zero particle state} \label{sec_distribution_MF_2}
Before we derive the evolution equation of the distribution function,
we discuss the specific representation of each wave function in the zero particle state.
The zero particle state $n_k^s=0$ corresponds to $E_k^s=\omega_k$ and $F_k^s=0$, that are equivalent to
\begin{eqnarray}
\omega_k&=&-s|{\boldsymbol k}|(|u_k^s|^2-|v_k^s|^2)
+sm_Ru_k^sv_k^{s*}+sm_R^*u_k^{s*}v_k^s, \nonumber \\ 
\\
0&=&-2s|{\boldsymbol k}|u_k^sv_k^s-sm_R(u_k^s)^2+sm_R^*(v_k^{s*})^2
\end{eqnarray}
from (\ref{eq:def_E}) and (\ref{eq:def_F}).
Solving the above equations about the wave functions $u_k^s$ and $v_k^s$,
one can obtain
\begin{eqnarray}
 u_k^s
  &=& e^{i\alpha}\cdot\sqrt{\frac{1}{2}\left(1-\frac{sk}{\omega_k}\right)}, \label{eq:zero_u}\\
 v_k^s
  &=& e^{i\alpha}\cdot\frac{sm_R}{|m_R|}\sqrt{\frac{1}{2}\left(1+\frac{sk}{\omega_k}\right)} \label{eq:zero_v}
\end{eqnarray}
where $\alpha$ is an arbitrary phase.
Note that the above results mean not functional solutions but just values at a time to be $n_k^s=0$.

\subsection{Equation of evolution} \label{sec_distribution_MF_3}
In this section, we consider obtaining the equation of evolution of the distribution function (\ref{eq:distribution_formula}).
Note that the original equations of motion of the wave functions (\ref{eq-EOMofMajo}) seem $2\times 2=4$ real degrees,
but the actual degrees are $4-1=3$ because there is a conservation law
\begin{equation}
 |u_k^s(t)|^2+|v_k^s(t)|^2=1.
\end{equation}
Since the distribution $n_k^s$ is described by $E_k^s$ that includes $|u_k^s|^2-|v_k^s|^2$ (1 real degree) and
$u_k^sv_k^{s*}$ (2 real degrees), it is enough to follow the evolution of these functions.
The time derivatives of these functions are given by
\begin{eqnarray}
 \partial_t(|u_k^s|^2-|v_k^s|)
  &=& 2is(m_Ru_k^sv_k^{s*}-m_R^*u_k^{s*}v_k^s), \\
 \partial_t[u_k^sv_k^{s*}]
  &=& ism_R^*(|u_k^s|^2-|v_k^s|^2)+2isku_k^sv_k^{s*}, \nonumber \\
\end{eqnarray}
that can be represented by a matrix form as
\begin{equation}
 \partial_t\left(\begin{array}{c}|u_k^s|^2-|v_k^s| \\ u_k^sv_k^{s*} \\ u_k^{s*}v_k^s \end{array}\right)
  =  isF \left(\begin{array}{c}|u_k^s|^2-|v_k^s| \\ u_k^sv_k^{s*} \\ u_k^{s*}v_k^s\end{array}\right) \label{eq:time_derivative_functions}
\end{equation}
\begin{equation}
 F \equiv \left(\begin{array}{ccc} 0 & 2m_R & -2m_R^* \\ m_R^* & 2k & 0 \\
  -m_R & 0 & -2k \end{array}\right).
\end{equation}
Eq.(\ref{eq:time_derivative_functions}) indicates the equations are closed within the three functions.
Because $E_k^s$ is described by the three functions $|u_k^s|^2-|v_k^s|^2,$ $u_k^sb_k^{s*}$, $u_k^{s*}b_k^s$,
the arbitrary higher derivations of $E_k^s$ also be described by those three functions.
This fact indicates that $E_k^s$, $\dot{E}_k^s$, $\ddot{E}_k^s$ can be independent, and thus
the higher derivatives $\partial_t^nE_k^s$ ($n\geq 3$) can be described by a linear combination of
$E_k^s$, $\dot{E}_k^s$, $\ddot{E}_k^s$.
Since the time derivatives of $E_k^s$ in (\ref{eq:def_E}) are given by
\begin{eqnarray}
 \dot{E}_k^s
  &=& s\dot{m}_R\cdot u_k^sv_k^{s*}+s\dot{m}_R\cdot u_k^{s*}v_k^s, \\
 \ddot{E}_k^s
  &=& i(m_R^*\dot{m}_R-\dot{m}_R^*m_R)(|u_k^s|^2-|v_k^s|^2) \nonumber \\
  & & +(s\ddot{m}_R+2ik\dot{m}_R)u_k^sv_k^{s*}+(s\ddot{m}^*_R*-2ik\dot{m}_R^*)u_k^{s*}v_k^s, \nonumber \\
\end{eqnarray}
one can obtain the matrix representation as
\begin{equation}
 \left(\begin{array}{c}E_k^s \\ \dot{E}_k^s \\ \ddot{E}_k^s \end{array}\right)
  = sG\left(\begin{array}{c}|u_k^s|^2-|v_k^s|^2 \\ u_k^sv_k^{s*} \\ u_k^{s*}v_k^s \end{array}\right) \label{eq:E_though_G}
\end{equation}
\begin{eqnarray}
 G &\equiv&
 \left(\begin{array}{ccc}-k & m_R & m_R^* \\ 0 & \dot{m}_R & \dot{m}_R^* \\
   is(m_R^*\dot{m}_R-\dot{m}_R^*m_R) & \ddot{m}_R+2isk\dot{m}_R & \ddot{m}_R^*-2isk\dot{m}_R^* \end{array}\right). \nonumber \\
\end{eqnarray}
Hence, the evolution of a set of $E_k^s$s are described by
\begin{eqnarray}
 \partial_t\left(\begin{array}{c}E_k^s \\ \dot{E}_k^s \\ \ddot{E}_k^s \end{array}\right)
  &=& s\dot{G} \left(\begin{array}{c}|u_k^s|^2-|v_k^s|^2 \\ u_k^sv_k^{s*} \\ u_k^{s*}v_k^s \end{array}\right)
   + iGF \left(\begin{array}{c}|u_k^s|^2-|v_k^s|^2 \\ u_k^sv_k^{s*} \\ u_k^{s*}v_k^s \end{array}\right) \nonumber \\\\
  &=& \left(\dot{G}+ isGF\right)G^{-1} \left(\begin{array}{c}E_k^s \\ \dot{E}_k^s \\ \ddot{E}_k^s \end{array}\right). \label{eq:eom_E_general}
\end{eqnarray}
In general, the components of $\left(\dot{G}+ isGF\right)G^{-1}$ are given by
\begin{equation}
 \left(\dot{G}+ isGF\right)G^{-1}
  = \left(\begin{array}{ccc} 0 & 1 & 0 \\ 0 & 0 & 1 \\ f_1 & f_2 & f_3 \end{array}\right)
\end{equation}
where $f_1, f_2, f_3$ are functions that consist of the time-dependent mass $m_R$, helicity $s$, and the momentum $|\boldsymbol k|$.

\vspace{2ex}

Let us consider a special case $\ddot{m}_R=0$.
Then one can obtain simple results:
\begin{equation}
 f_1=4{\rm Re}[m_R^*\dot{m}_R], \quad f_2=-4\omega_k^2, \quad f_3=0,
\end{equation}
that lead
\begin{eqnarray}
 \dddot{E}_k^s
  &=& f_1E_k^s+f_2\dot{E}_k^s+f_2\ddot{E}_k^s \nonumber \\
  &=& 4{\rm Re}[m_R^*\dot{m}_R]\cdot E_k^s-4\omega_k^2\dot{E}_k^s.
\end{eqnarray}
Choosing the initial condition to be the zero particle state,
each of the initial values is given by
\begin{equation}
 \left(\begin{array}{c}E_k^s \\ \dot{E}_k^s \\ \ddot{E}_k^s \end{array}\right)
  = \left(\begin{array}{c} \omega_k \\ {\rm Re}[m_R^*\dot{m}_R]/\omega_k \\ 0 \end{array}\right)
\end{equation}
where we used (\ref{eq:zero_u}), (\ref{eq:zero_v}), and (\ref{eq:E_though_G}).
Note in this case that $E_k^s$ can evolve but $E_k^+-E_k^-$ cannot because
the equation of evolution is described by
\begin{eqnarray}
 \partial_t^3(E_k^+-E_k^-)
  &=& 4{\rm Re}[m_R^*\dot{m}_R]\cdot (E_k^+-E_k^-) \nonumber \\
  & & -4\omega_k^2\cdot\partial_t(E_k^+-E_k^-)
\end{eqnarray}
and all the initial values of $E_k^+-E_k^-$, $\partial_t(E_k^+-E_k^-)$, $\partial_t^2(E_k^+-E_k^-)$ are zero.
Thus, we can conclude that the net number
\begin{equation}
 n_k^+-n_k^-=-\frac{E_k^+-E_k^-}{2\omega_k}
\end{equation}
also never develops from the zero particle state in the case of $\ddot{m}_R=0$.
One can also obtain the same statement in the case of $\boldsymbol k=0$.

\end{document}